%% file: MTD_CCPA_journal_R2_revision.tex
\documentclass[journal]{IEEEtran}
\usepackage{graphicx,epsfig,amsmath,amssymb,bm}
\usepackage{amssymb,balance}
\usepackage{amsmath}
\usepackage{amsthm} 
\usepackage{tikz}
\usepackage{amssymb}
\usepackage{algorithmicx}
\usepackage{algpseudocode}
\usepackage{url}
\input{macros}

\usepackage{bbm}
\usepackage{enumitem}
\usepackage{subcaption}
\usepackage{tabularx}
\usepackage{multirow}
\usepackage{verbatim}
\usepackage[ruled,linesnumbered]{algorithm2e}

\newtheorem{proposition}{Proposition}

\newtheorem{problem}{Problem}

\newtheorem{definition}[theorem]{Definition}
\newcommand{\beqa}{\begin{eqnarray}}
\newcommand{\eeqa}{\end{eqnarray}}
\newcommand{\dsp}{\displaystyle}

\ifCLASSINFOpdf

\else

\fi

\graphicspath{{/Users/subhash/Dropbox/ADSC/Since_2017/Latex/MTD/Figures/}}

\begin{document}

\title{Moving-Target Defense Against Cyber-Physical Attacks in Power Grids via Game Theory}
\author{Subhash~Lakshminarayana,~\IEEEmembership{Senior Member, IEEE}, E. Veronica Belmega,~\IEEEmembership{Senior Member, IEEE},\\ and H. Vincent Poor,~\IEEEmembership{Fellow, IEEE} \vspace{-0.25in}
\thanks{S. Lakshminarayana is with the University of Warwick, Coventry, UK (subhash.lakshminarayana@warwick.ac.uk). E.V. Belmega is
with ETIS, UMR 8051, CY Cergy Paris Université, ENSEA, CNRS, F-95000, France (belmega@ensea.fr). H. V. Poor is with the Department of Electrical Engineering, Princeton University, USA (poor@princeton.edu). The work was partially presented at IEEE Smartgridcomm-2019 \cite{LakshCCPA2019}. This research was supported in part by a Startup grant at the University of Warwick, Cergy : ENSEA (SRV), Paris Seine EUTOPIA international fellowship, COST Action CA16228 “European Network for Game Theory” (GAMENET),  and by the U.S. National Science Foundation under Grants  ECCS-1824710 and ECCS-203971.
}}



\maketitle

\begin{abstract}
This work proposes a moving target defense (MTD) strategy to detect coordinated cyber-physical attacks (CCPAs) against power grids. The main idea of the proposed approach is to invalidate the knowledge that the attackers use to mask the effects of their physical attack by actively perturbing the grid's transmission line reactances via distributed flexible AC transmission system (D-FACTS) devices. 
The proposed MTD design consists of two parts. First, we identify the subset of links for D-FACTS device deployment that enables the defender to detect CCPAs against any link in the system. Then, in order to minimize the defense cost during the system's operational time, we formulate a zero-sum game to identify the best subset of links 
to perturb  (which will provide adequate protection) against a strategic attacker. The Nash equilibrium robust solution is computed via exponential weights, which does not require complete knowledge of the game but only the observed payoff at each iteration.
Extensive simulations performed using the MATPOWER simulator on IEEE bus systems verify the effectiveness of our approach in detecting CCPAs and reducing the operator's defense cost. 
\end{abstract}
{\IEEEkeywords moving-target defense, coordinated cyber-physical attacks,  zero-sum non-cooperative games, Nash equilibrium, exponential weights algorithm}

\IEEEpeerreviewmaketitle

\section{Introduction}
Cyber threats against power grids are of increasing concern due to the deep integration of information and communication technologies (ICT) into grid operations. In particular, the so-called coordinated cyber-physical attacks (CCPAs) can be dangerously covert. Indeed, while the physical attack involves disconnecting a transmission line, generator or transformer, the simultaneous cyber attack plays the role of masking the physical attack by manipulating the sensor measurements that are conveyed from the field devices to the control center. CCPAs can have severe effects on the grid, since undetected line/generator outages may trigger cascading failures, and have received significant attention \cite{SoltanCCPA2015, LiCCPA2016, LiCCPA2018, DengCCPA2017}.

To defend against CCPAs, recent studies \cite{LiCCPA2016} and \cite{DengCCPA2017} have proposed  strategies based either on securing a set of measurements (e.g., by encryption) or relying on measurements from known-secure phasor measurement units (PMU) deployed in the grid. However, power grids consist of many legacy devices whose life cycles can last several decades, and incorporating major security upgrades in these devices can be quite expensive. Moreover, extensive research has shown that PMUs themselves are vulnerable to false data injection (FDI) attacks, which can be launched by spoofing their GPS receivers \cite{ShepardGPS2012}. For a general class for false data injection attacks against state estimation, recent works \cite{OzayML2016, HeDL2017} have proposed machine learning (ML) based methods to detect the attacks. References \cite{HabibiDC2020, HabibiDCtwo2020} propose ML-based defense for FDI attacks against DC micro-grids. However, recent research has shown that ML-based algorithms can be vulnerable in adversarial scenarios, which can significantly reduce their efficacy \cite{SaygheISGT2020}. Thus, existing defense mechanisms are not foolproof.

In this paper, we propose a novel defense strategy to detect CCPAs based on the moving target defense (MTD) technique. As opposed to traditional static defense, MTD is a dynamic strategy that has the potential to increase the complexity and the cost for potential attackers. 
As in prior works \cite{SoltanCCPA2015, LiCCPA2016, LiCCPA2018, DengCCPA2017}, we consider physical attacks that disconnect transmission lines. We note that to craft an undetectable CCPA, the attacker must obtain an accurate knowledge of the transmission line reactances \cite{LiCCPA2016, DengCCPA2017}. The main idea of the proposed MTD defense in this context is to invalidate the attacker's prior acquired knowledge by actively perturbing of the grid's line reactance settings. This can be accomplished using distributed flexible AC transmission system (D-FACTS) devices, which are capable of performing active impedance injection and are being increasingly deployed in power grids  \cite{DFACTS2007}. The proposed MTD defense strategy has the potential to make it extremely difficult for the attacker to track the system's dynamics and gather sufficient information to craft {covert} CCPA. 


\paragraph*{{Recent related works}} \cite{LakshDSN2018, LiuMTD2018, TianHiddenMTD2019, ZhangMTD2020} have studied MTD to defend the power grid's state estimation against false data injection (FDI) attacks.
References \cite{LakshDSN2018} and \cite{LiuMTD2018} proposed a design criteria to compute MTD perturbations that can detect FDI attacks effectively. They also showed that effective MTD perturbations incur a non-trivial operational cost, evaluated in terms of the grid's optimal power flow (OPF) cost \cite{LakshDSN2018} and the transmission power losses \cite{LiuMTD2018}. The problem of hiding MTD's activation from the attacker was considered in \cite{TianHiddenMTD2019}, which proposed the so-called \emph{hidden MTD}. Reference \cite{ZhangMTD2020} analyzed how the topology of the power grid impacts the completeness of MTD’s detection capability. Finally, \cite{LiuPlacement2020}
has studied the deployment of D-FACTS devices to maximize MTD's attack detection capability against FDI attacks.

Compared to the above references, {the novelty of our work is two-fold.} First, with the exception of our preliminary work \cite{LakshCCPA2019}, none of them have considered defense against CCPAs by exploiting MTD. The solution requires the formulation of novel design criteria both in terms of D-FACTS placement as well as D-FACTS perturbation selection. Second, existing works seek to design MTD that can defend against all potential threats (specifically, FDI attacks in these papers). However, as we will show in this work, this is not always necessary. In fact, MTD's operational cost can be significantly reduced by identifying and defending against only the most likely threats against an active and strategic attacker. We formalize this idea in the context of MTD for CCPAs using a game-theoretic framework. We note that although game theory has been used in the context of defense against FDI attacks \cite{EsmaTSG2013, SanjabTSG2016}, our work is the first to apply it in the context of MTD.

\paragraph*{Our main contributions} in the MTD design against CCPAs concern two different aspects: (i) \emph{D-FACTS deployment}, and (ii) \emph{D-FACTS operation}. First, in the \emph{D-FACTS deployment} problem, we seek to find the best subset of links for this purpose that combines the dual criteria of minimizing the system's OPF cost and detecting CCPAs. We identify a graph-theoretic criteria to characterize MTD's effectiveness against CCPAs. The optimal subset of links for D-FACTS deployment are found by solving a feedback edge set problem \cite{BondyGraph1976} on the graph associated with the power grid. Our proposed D-FACTS deployment solution provides the defender with the ability to detect CCPAs against any transmission line of the grid.

Second, to reduce the prohibitive MTD operational cost \cite{LakshDSN2018}, we focus on \emph{D-FACTS operation} that minimizes the defense cost while accounting for a strategic and intelligent attacker.\footnote{A key observation that enables the cost reduction is that, during system operations, it is not necessary to defend against CCPAs all transmission lines in the grid. This is because many of these attacks are not harmful enough and, hence, the attacker is unlikely to target those transmission lines.} For this, we study the interaction between the grid's defender and attacker via zero-sum non-cooperative games. This enables us to anticipate the attacker's strategic behaviour and to develop robust defense policies against CCPAs. Then, in order to compute the Nash equilibrium (NE) minimax robust solution, we propose to exploit a reinforcement learning algorithm, namely, the exponential weights or EXP3 \cite{auer-2003}. The latter has two major advantages compared to the Lemke-Howson algorithm: (1)  EXP3 does not require the perfect and complete knowledge of the game, which may be problematic in adversarial settings, and relies  solely on the past observed payoff; (2) Lemke-Howson is a combinatorial algorithm and, although efficient in practice, it has exponential complexity in the worst case. At the opposite, EXP3 is a simple iterative procedure converging to the solution as $\mathcal{O}({T}^{-1/2})$, with $T$ being the horizon of play.

At last, our extensive simulations conducted using the MATPOWER simulator shows the effectiveness of the proposed MTD solution in detecting CCPAs. Our results also demonstrate that the game-theoretic robust solution significantly reduces the operator's defense cost. Moreover, the simulation results also demonstrate that MTD designed according to the DC power flow model remains effective in detecting CCPA attacks under an AC-power flow model as well.

Compared to our \emph{preliminary work \cite{LakshCCPA2019}}, the novel contributions of this paper are significant, as follows: (i) we provide a new D-FACTS deployment scheme that in addition to defending against CCPAs (as in \cite{LakshCCPA2019}) also minimizes the OPF cost simultaneously; (ii) EXP3 is proposed to compute the NE instead of the Lemke-Howson approach in \cite{LakshCCPA2019}, which requires less information and comes with performance guarantees; (iii) multi-line attacks are considered here, while only single line disconnections were considered in \cite{LakshCCPA2019};  (v) as opposed to the full placement of power flow/injection measurement sensors at all links/busses assumed in \cite{LakshCCPA2019}, here we also consider partial sensor placement; and (iv) we show via numerical experiments that our defense strategy remains effective in the AC-power flow case.

Finally, we note that although in this paper, we focus specifically on strengthening the BDD to detect CCPAs, the proposed MTD method can be broadly applied to other data-driven methods for line outage detection/classification such as those proposed in \cite{TateOverbye2008, HaoLineOutage2012,  VeerTSP2017, ZhaoLineoutage2020}. These data-driven approaches leveraging the measurements provided by the supervisory control and data acquisition (SCADA) system and PMUs are being increasingly adopted in power grids, and an attacker aiming to cause line outages must mask the effect of the physical attack on the power grid measurements to remain stealthy. Thus, a similar MTD strategy can also be developed to strengthen the aforementioned data-driven line outage detection/classification approaches.


\section{System Model}
\label{sec:Prelim}

We consider a power grid characterized by a graph $\mathcal{G = (N,L)},$
where $\mathcal{N} = \{1,\dots,N\}$ is the set of buses and $\mathcal{L} = \{1,\dots,L\}$ is the set of transmission lines. Without the loss of generality, we assume throughout that bus $1$ is the slack bus.
At bus $i,$ we denote the amount of generation and load by
$G_{i}$ and $L_{i}$ respectively.  We let $l = \{ i,j\}$ denote a transmission line $l \in \mathcal{L}$ 
that connects bus $i$ and bus $j$ and its reactance by $x_{l}.$
The power flowing on the corresponding line $l$ is denoted by $F_{l},$ which under the DC power flow model 
is given by $F_{l} = \frac{1}{x_{l}}(\theta_{i} - \theta_{j}),$ where $\theta_{i}$ and $\theta_{j}$ are the voltage phase angles at buses $i,j \in \mathcal{N}$ respectively. Note for the slack bus $\theta_1 = 0$. In vector form, the power flow vector  $\fv = [F_{1},\dots,F_{L}]^T$ is related to the voltage phase angle vector  $\thetav = [ \theta_2,\dots,\theta_N] $ as  $\fv = \Dm \Am^T \thetav,$ where the matrix $\Am \in \RR^{N-1 \times L}$ is the  reduced branch-bus incidence matrix obtained by removing the row of the slack bus and $\Dm \in \RR^{L \times L}$ is a diagonal matrix of the reciprocals of link reactances.
We denote the set of links on which D-FACTS devices are deployed by $\mathcal{L}_D$ where $\mathcal{L}_D \subseteq \mathcal{L}.$ D-FACTS devices enable the reactances of these lines to be varied within a pre-defined range $[\xv^{\min} , \xv^{\max}],$ where $\xv^{\min},\xv^{\max}$ are the reactance limits achievable by the D-FACTS devices. Note that 
$x^{\min}_l = x^{\max}_l = x_l, \forall l \in \mathcal{L} \setminus \mathcal{L}_D.$


\subsubsection*{State Estimation \& Bad Data Detection}
The system state, i.e., the voltage phase angles $\thetav,$ are estimated from the noisy sensor measurements using the state estimation (SE) technique. The sensor measurements, which we denote by ${\zv} \in \RR^{M },$ correspond to the 
nodal power injections, and the forward and reverse branch power flows, i.e. ${\zv} = [\tilde{\pv},\tilde{\fv},-\tilde{\fv}]^T$ and $M$ is the total number of measurements\footnote{ { We will generealize our results to the case of partial placement of sensors in Sec.~\ref{sec:MTD_Soln}-B.}}, where $M = N+2L.$ We denote the sensor measurement noises by a vector $\nv \in \RR^{M },$ which is assumed to follow a Gaussian distribution.
Under the DC power flow model, the relationship between $\thetav$ and $\zv$ is given by 
${\zv} = \Hm \thetav + \nv,$ where $\Hm \in \RR^{M \times N}$ is the system's measurement matrix given by
$\Hm = [\Dm \Am^T;-\Dm \Am^T;\Am \Dm \Am^T]$.
The maximum likelihood (ML) technique is used for system state estimation. 
Under ML estimation, the estimate $\widehat{\thetav}$ is related to the measurements $\zv$ as
$\widehat{\thetav} = ({\Hm^T} \Wm \Hm)^{-1} {\Hm^T} \Wm \zv,$
where $\Wm$ is a diagonal weighting matrix whose elements are reciprocals
of the variances of the sensor measurement noise components.

After state estimation, a bad data detector (BDD) computes the residual, which we denote by $r$, as $r = ||\zv - \Hm \widehat{\thetav}||.$ A bad data alarm is flagged if the residual exceeds a predefined threshold  $\tau.$ The threshold is adjusted to ensure that the false positive (FP) rate
does not exceed $\alpha,$ where $\alpha > 0$ (usually a small value close to zero).

\subsubsection*{Optimal Power Flow}
For any given load condition $\dv = [D_1,\dots,D_N]$, the system operator sets the generation dispatch and line reactance settings by solving the optimal power flow (OPF) problem, 
stated as follows:
\begin{subequations}
\label{eqn:OPF_normal}
\beqa
 C_{\text{OPF}}  =  & \dsp \min_{ \pv_g, \qv_g,\xv} &  \sum_{i \in \mathcal{N}} C_i (P_{G_i}) \label{eqn:OPF_normala}
    \\ 
& s.t. &  \gv - \dv  = \Bm \thetav, \label{eqn:OPF_normalb}
 \\
 & & | \gv - \gv^{\prime} | \leq \Delta^{\max}_g \label{eqn:OPF_normalc}\\
& & \fv \in \mathcal{F}, \gv \in \mathcal{Z}, \xv \in \mathcal{X},  \label{eqn:OPF_normald}
\eeqa 
\end{subequations}
where $C_i(\cdot)$ is the generation cost 
at bus $i \in \mathcal{N}$ and $\pv_g = [P_{G_1},\dots,P_{G_N}]$ is the power generation vector at all busses. Equation \eqref{eqn:OPF_normalb} is the nodal power balance constraint, where the matrix $\Bm =  \Am \Dm \Am^T.$  Constraint \eqref{eqn:OPF_normalc} is the generator ramp rate constraint, where $\gv^\prime = [G^\prime_1,\dots,G^\prime_N]$ is the vector of generations at the previous decision instant and $\Delta^{\max}_g$ is the permissible change in the generation between two decision instants.  Constraints \eqref{eqn:OPF_normald} correspond to the branch power flows, generator limits, and D-FACTS limits, respectively, where $\mathcal{F} = [-\fv^{\max},\fv^{\max}], \mathcal{Z} = [\gv^{\min},\gv^{\max}]$ and $\mathcal{X} = [\xv^{\min}, \xv^{\max}]$ and $\fv^{\max}$ is the maximum permissible line power flow (i.e., the thermal limit) and $\gv^{\min},\gv^{\max}$ are the generator limits. We note that in the absence of D-FACTS, OPF optimizes over the generator dispatch values only.

\section{Coordinated Cyber and Physical Attacks}
\label{sec:CCPA}
The focus of this work is the power grid SCADA system. Existing SCADA communication standards, such as the IEEE C37.118 and IEC-61850 frameworks  are known to have poor security features (e.g., lack of encryption, etc.) \cite{PMUComm2016}. Their vulnerabilities can be exploited by an attacker to obtain unauthorized access to the field devices and alter their status, and/or modify the data packets which convey the field measurements and the control commands.

\paragraph*{Undetectable False Data Injection Attacks}
We denote the false data injection (FDI) attack vector by $\av \in \RR^{M},$ which the attacker injects into the sensor measurements by exploiting the aformentioned SCADA vulnerabilities. The measurement vector with the FDI attack by $\zv^a$, given by $\zv^a = \zv+\av.$ It has been shown 
\cite{Liu2009} that an FDI attack of the form $\av = \Hm \cv,$ where $\cv \in \RR^N,$ remains undetected by the BDD.  Specifically, the probability of detection for such attacks is equal to the FP rate  $\alpha.$ We call these attacks undetectable FDI attacks. 

\paragraph*{Coordinated Cyber and Physical Attack}
While an FDI attack only modifies the sensor measurements, a CCPA harms the grid physically followed by a coordinated FDI attack on the sensor measurements, mentioned above. In particular, we consider physical attacks that disconnect a set of transmission lines, e.g., by opening the line circuit breakers by accessing them remotely exploiting the SCADA vulnerabilities. The physical attack will alter the power grid's topology and power flow, and the mismatch between the pre-attack (i.e., line disconnections) and post-attack measurements can generally be detected by the BDD. However, it has been shown that if the attacker injects a carefully-constructed coordinated FDI  attack on the sensor 
measurements, then the effect of the physical attack on the BDD residual can be completely masked \cite{DengCCPA2017}. Hence, the attack remains undetected by the BDD. 

Denote the set of links disconnected by the attacker under a physical attack by $\mathcal{L}_A.$
We use the subscript $``p"$ to denote the power grid parameters following the physical attack. It can be shown that the grid measurements post physical attack are related to the pre-attack measurements by
 $\zv_p = \zv+\av_p,$ where $\av_p = \Hm \Delta \theta + \Delta \Hm \thetav_p,$ where $ \Delta \Hm$ is the change in the measurement matrix before and after the physical attack, given by, $\Delta \Hm = \Hm-\Hm_p.$ As shown in \cite{DengCCPA2017}, in order to mask the effect of the physical attack and remain undetected by the BDD, the attacker must inject a coordinated FDI attack of the form $\av = \Delta \Hm \thetav_p.$

\begin{figure}[!t]
\centering
\includegraphics[width=0.4\textwidth]{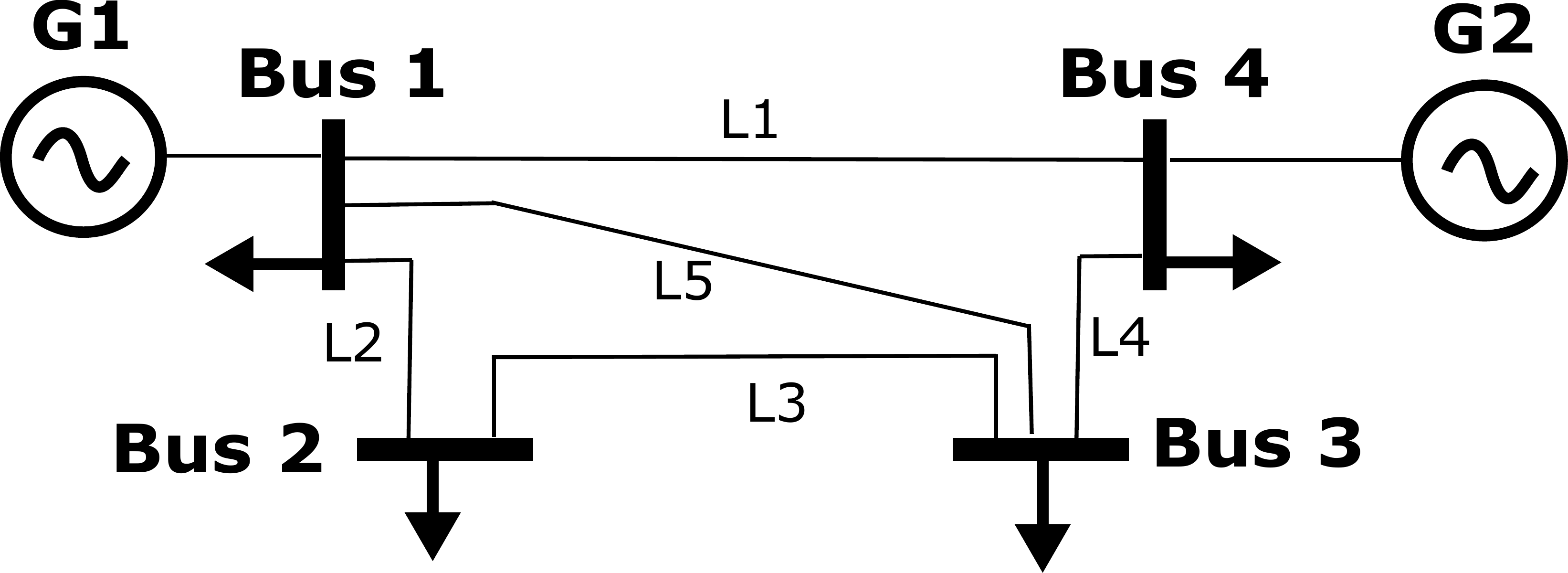}
\caption{An example of a $4$ bus power grid. }
\label{fig:4bus}
\vspace{-0.4 cm}
\end{figure}

\paragraph*{Knowledge Required to Launch a CCPA}
Next, we enlist the knowledge required by the attacker to construct an FDI attack of the form $\av = \Delta \Hm \thetav_p.$ Assume that the attacker disconnects a single branch $\mathcal{L}_A = \{l \}$ that connects buses $i$ and $j.$ 
It can be easily verified that $\Delta \Hm$ depends on the tripped branch 
 reactance $x_l$ only. Therefore, to construct the attack $\av = \Delta \Hm \thetav_p$, the attacker must obtain knowledge of the branch reactance $x_l$ and the difference in phase angles of the buses $i$ and $j$ following the physical attack, i.e., $\theta_{i,p} - \theta_{j,p}$ \cite{DengCCPA2017}. The knowledge of $\theta_{i,p} - \theta_{j,p}$ can be obtained by monitoring the line power flows following the physical attack as follows:
\begin{align}
 \theta_{i,p} - \theta_{j,p} = - \sum_{ m \in p^{k}_l }  x_{l_m} F_{l_m,p}, \label{eqn:ph_diff}
\end{align}
where $p^k_{l} $ is any alternative path between nodes $i$ and $j$ in the residual power network following the physical disconnections, i.e., $\mathcal{L} \setminus \mathcal{L}_A. $ Each path $p^k_{l}$ in turn is a collection of links $p^k_{l} = \{ l_{k_1}, l_{k_2}, \dots, l_{k_M}\}$ such that
$src(l_{k_1}) = i$ and $dst(l_{k_M}) = j,$ and $k_M$ is the number of links in the path  $p^k_{l}.$
We denote by $\mathcal{P}_{l} = \{ p^1_{l}, p^2_{l}, \dots, p^{K_l}_{l} \}$ a collection of all alternative paths between buses $i$ and $j,$ where $K_l$ is the number of such alternative paths. Note that the subscript $l$ denotes the disconnected link.

In the IEEE-4 bus example shown in Fig.~\ref{fig:4bus}, assume that the attacker disconnects link~1. 
After the disconnection, there are two alternative paths between buses $1$ and $4,$ and hence, $K_1 = 2.$ These paths are given by
$p^{1}_l = \{ 2,3,4 \}$ with $k_1 = 3$ and  $p^{2}_l = \{ 5,4 \}$ with $k_2 = 2.$ The attacker can compute the phase angle difference between nodes 1 and 2 using \eqref{eqn:ph_diff} as
$\theta_{1,p} - \theta_{j2p}  = - \LB x_2 F_{2,p} + x_3 F_{3,p} + x_4 F_{4,p} \RB$ or,  $ \theta_{1,p} - \theta_{j2p}  = - \LB x_5 F_{5,p} + x_4 F_{4,p} \RB.$

The attacker can obtain the knowledge of power flows $F_{l_m,p}$ in \eqref{eqn:ph_diff} by monitoring the line flow sensor measurements. The line reactances $x_{l_m}$ can be learned by monitoring the grid power flows over a period of time using existing techniques \cite{LakshDataDrive2020}. The attacker can also learn the reactance of the disconnected branch $x_l$ similarly.

\paragraph*{CCPA with Multiple Line Disconnections} While the description above considers CCPAs with single line connections, an attacker who has access to the line circuit breakers can simultaneously disconnect multiple tranmission lines. In this case, to launch an undetectable CCPA, the attacker would need to acquire the knowledge of the reactances of all tripped branches, and the phase angles of all buses connecting to the tripped branches after the physical attack \cite{DengCCPA2017}.

\section{Moving-Target Defense for CCPAs}
\label{sec:MTD}
In this work, we propose a solution to defend the system against CCPAs based on the MTD technique.  In the following, we first describe MTD for power grids and then formalize the MTD design problem to defend against CCPAs.

\subsection{ MTD Description and Practical Implementation}
The main idea behind this approach is to periodically perturb the branch reactances of certain transmission lines to invalidate the attacker's acquired knowledge. Hence, an attack constructed using outdated knowledge of the system can be detected by the BDD. Two important considerations in MTD design are the MTD cost and the perturbation frequency, which we explain in the following. 

\emph{ MTD Cost:} Note that in the absence of MTD consideration, the D-FACTS device settings are optimized to minimize the OPF cost as in \eqref{eqn:OPF_normal}. Let us denote $\gv^*, \xv^* \defines \arg \min_{\gv,\xv}  C_{\text{OPF}} .$  However, under MTD, to invalidate the attacker's knowledge, the reactance settings are set to be different from $\xv^*.$ Thus, MTD perturbations will incur an operational cost \cite{LakshDSN2018}.

Let us denote the line reactnace settings with MTD as $\xv^\prime = \xv + \Delta \xv,$ where $\Delta \xv$ is the reactance perturbation. The OPF cost with MTD is given by
\begin{subequations}
\label{eqn:OPF_MTD}
\beqa
 C^\prime_{\text{OPF}}  =  & \dsp \min_{ \gv} &  \sum_{i \in \mathcal{N}} C_i (G_{i}) \label{eqn:OPF_MTDa}
    \\ 
& s.t. &  \gv - \lv  = \Bm \thetav, \label{eqn:OPF_MTDb}
 \\
 & & | \gv - \gv^{\prime} | \leq \Delta^{\max}_g \label{eqn:OPF_MTDc}\\
 & & \xv =  \xv^* + \Delta \xv \\
& & \fv \in \mathcal{F}, \gv \in \mathcal{Z}.   \label{eqn:OPF_MTDd}
\eeqa 
\end{subequations}
The cost of MTD is the difference between the OPF cost without MTD and OPF cost with MTD, i.e., $C_{\text{MTD}} = C_{\text{OPF}} - C^\prime_{\text{OPF}}.$ Note that the MTD cost is always non-negative since the additional perturbation due to MTD will increase the OPF cost. Moreover, the magnitude of $\Delta \xv$ affects the MTD cost. In general, a large perturbation effectively invalidates the attacker's knowledge, but also incurs higher operation cost. Thus, there exists a trade-off between MTD's effectiveness and cost \cite{LakshDSN2018}.

\emph{MTD Perturbation Frequency:} For MTD to be effective, the system settings must be changed before the attacker gathers sufficient information to conduct a successful attack. The attacker can acquire the knowledge of the system topology and branch reactances (required to launch undetectable CCPA) by monitoring the power grid's measurement data over a period of time \cite{LakshDataDrive2020}. The experimental evidence suggests that hourly perturbations are sufficient for practical systems (see the discussion in Section IV of \cite{LakshDSN2018}).

\begin{figure}[!t]
\centering
\includegraphics[width=0.48\textwidth]{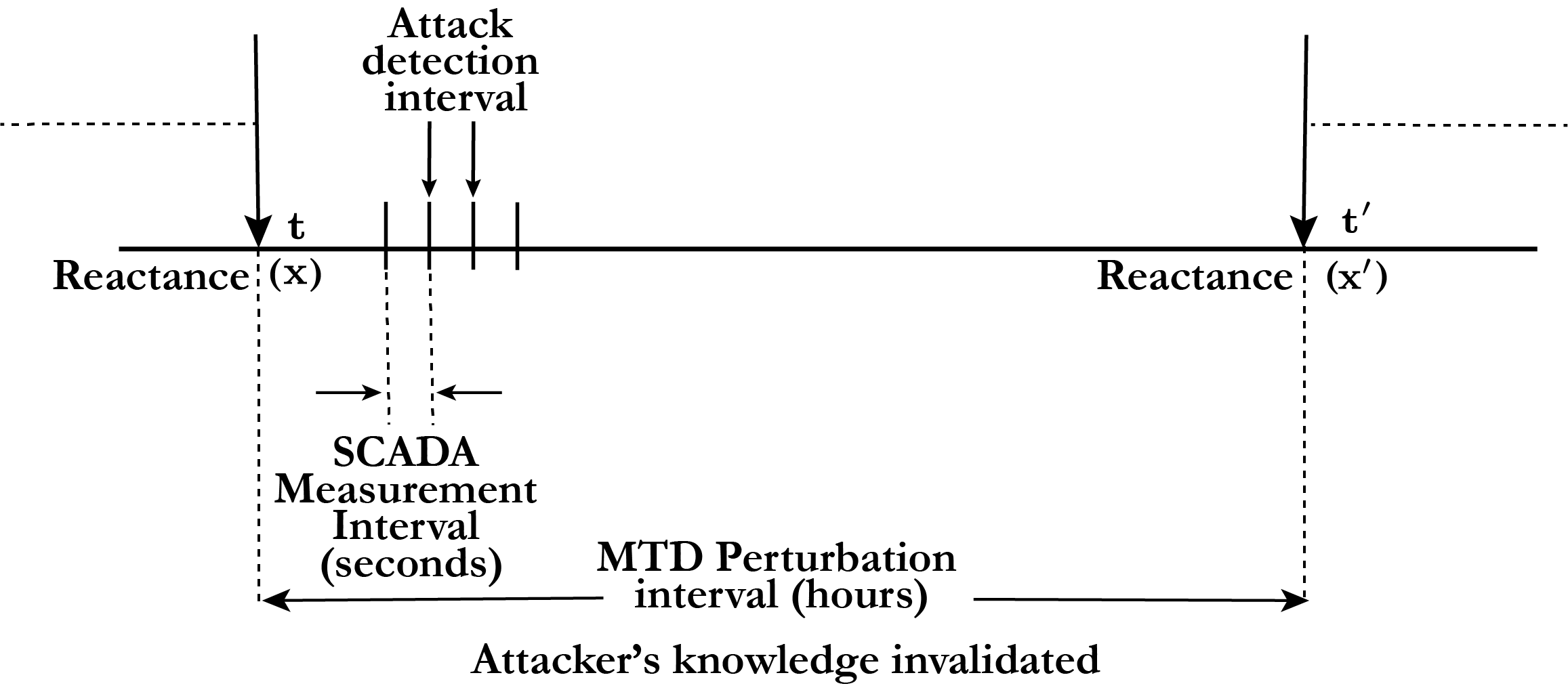}
\caption{A time line of the proposed MTD scheme.}
\label{fig:Timeline}
\vspace{-0.4 cm}
\end{figure}

\emph{Practical Implementation:} A time line showing the practical implementation of the proposed MTD scheme is presented in Fig. \ref{fig:Timeline}. As shown in the figure, the MTD perturbation interval is on the order of hours, whereas the SCADA measurement frequency is in the order of 4-6 seconds (considering traditional SCADA) and about 50 measurements per second considering the advanced PMUs. 
When the reactance of the transmission lines is perturbed, the attacker's acquired knowledge of the power system becomes invalid. Thus, any CCPA executed by the attacker with an outdated knowledge of the system will be detected by the BDD, once the next set of field measurements from the SCADA system reach the control center. Thus, BDD strengthened with the MTD will be able to detect CCPAs within this time frame of the SCADA's measurement frequency.

\subsection{Defending Against CCPAs}
Next, we formalize the MTD design problem to defend against CCPAs. For clarity, we set up the problem for CCPAs involving single line disconnections only. However, in Section V-A, we present arguments to show that the proposed MTD design remains effective against CCPAs with multiple line disconnections. 

Recall from \eqref{eqn:ph_diff} that to construct an undetectable CCPA, the attacker must acquire the following: (i) knowledge of the reactance of the tripped branch, $x_l,$ and (ii) knowledge of branch reactances in at-least  one alternate paths $p^k_{l}$
between the nodes $i$ and $j.$ 
Therefore, under MTD, the defender can thwart the CCPA by invalidating one of the two:
\begin{itemize}
\item[C1.]  Invalidate the attacker's knowledge of the tripped branch's reactance $x_l$. 
\item[C2.]  Invalidate the attacker's knowledge of at-least one of the branches in the path $p^k_{l}$ between nodes $i$ and $j.$
\end{itemize}
Note that the defender cannot have prior knowledge of which link the attacker chooses to disconnect. Moreover, for a disconnected link 
$l,$ the defender has no way of knowing which path $p^k_{l} \in \mathcal{P}_{l}$ the attacker may have used to compute the phase angle difference $\theta_{i,p} - \theta_{j,p}$ as in \eqref{eqn:ph_diff}. Thus, the defender must invalidate the attacker's knowledge of the reactance of at-least one link in every path $p^k_{l} \in \mathcal{P}_{l}.$ The defender must do so for every link $l \in \mathcal{L}$ (such that the attacker cannot launch a CCPA by disconnecting any link in the grid). 

To sum up, the MTD perturbation selection problem can be stated as follows:
\begin{problem}[MTD design]
For each branch $l \in \mathcal{L},$ invalidate the knowledge of at-least one of the branches in $ \{ l \} \cup p^k_{l}, \ k = 1,\dots, K_l.$
\end{problem}
This problem poses constraints on the D-FACTS deployment set  $\mathcal{L}_D,$ since a preliminary requirement to invalidate the attacker's knowledge of a branch reactance is the presence of a D-FACTS device on that link. Thus, $\mathcal{L}_D$ must be chosen in a way that it gives the defender the ability to protect every link $l \in \mathcal{L}$. A trivial solution is to deploy a D-FACTS device on every link of the power grid. However, a system operator may wish to minimize the number of D-FACTS devices installed in order to minimize the device deployment cost. 



On the other hand, MTD perturbations also incur an operational cost for the defender as shown in \cite{LakshDSN2018}.
Indeed, perturbing the reactances of a large number of links may be expensive. Thus, at the system's operational time, the defender may wish to perturb the reactances of only a subset of links, which we denote by  $\mathcal{L}_{D_w},$ where  $\mathcal{L}_{D_w} \subseteq \mathcal{L}_D,$ such that the attacker cannot launch CCPAs against some specific links that are perceived to be critical and vulnerable.


In what follows, we provide solutions to both the aforementioned aspects of MTD design problem. 
Specifically, in Section~\ref{sec:MTD_Soln}, we first present an algorithm to find the D-FACTS deployment set $\mathcal{L}_D$ that satisfies the MTD design problem with a minimum number of devices based on a graph-theoretic approach. Subsequently, in Section~\ref{sec:Game}, we present a solution to the problem of selecting a subset of links $\mathcal{L}_{D_w}$ for 
reactance perturbation at the operational time based on a game-theoretic approach.

\section{D-FACTS Deployment Solution}
\label{sec:MTD_Soln}
In the absence of defense consideration, D-FACTS are traditionally deployed in the grid with an objective of minimizing the cost of operation \cite{RogersDFACTS2008}. In this work, we seek to find the D-FACTS deployment set
$\mathcal{L}_D$ that balances the dual objectives of defense against CCPAs and minimizing the OPF cost. Specifically, $\mathcal{L}_D$ must satisfy the following criteria:
\emph{(i)} $\mathcal{L}_D$ must meet the CCPA detection conditions listed in Section~\ref{sec:MTD} and the number of links in $\mathcal{L}_D$, denoted by $|\mathcal{L}_D|$, must be minimal; and \emph{(ii)} $\mathcal{L}_D$ must achieve the least OPF cost. 

First, we consider the problem of finding $\mathcal{L}_D$ that satisfies \emph{(i)} above. For this, the key observation is that each set of links $\{l \} \cup p^k_l, k = 1,\dots, k_M,$ forms a loop in the graph $\mathcal{G}.$ In Figure~\ref{fig:4bus}, assuming that the attacker disconnects link 1, the links $\{1 \} \cup \{2,3,4 \} $ and $\{1 \} \cup \{4,5 \} $ form loops in the corresponding graph. If D-FACTS devices are installed on a subset of links in the graph such that every loop contains at least one D-FACTS link, then the attacker cannot launch an undetectable CCPA. In graph-theoretic terms, the problem is equivalent to removing a subset of links in the network such that the residual graph has no loops. For optimized deployment, $\mathcal{L}_D$ must contain the minimum number of links. 

The set $\mathcal{L}_D$ can be found by solving the \emph{feedback edge set problem in an undirected graph} \cite{BondyGraph1976}. The solution proceeds by finding the \emph{spanning trees} of the graph $\mathcal{G}$. Let $\mathcal{L}_{\text{sptr}}$ denote a spanning tree of $\mathcal{G}.$ If D-FACTS devices are installed on the links $\mathcal{L} \setminus \mathcal{L}_{\text{sptr}},$ then the attacker cannot find a loop within the graph whose branches do not have a D-FACTS device. Equivalently, the attacker cannot launch an undetectable CCPA. 
Further, by definition, every spanning tree in a connected graph contain precisely $N-1$ links and, hence, $\mathcal{L} \setminus \mathcal{L}_{\text{sptr}}$ contains the minimum number of links that can be disconnected and satisfy \emph{(i)}. 
Thus, the D-FACTS deployment set is $\mathcal{L}_D = \mathcal{L} \setminus \mathcal{L}_{\text{sptr}}.$ 


The graph $\mathcal{G}$ might have multiple spanning trees, which imply multiple subsets of links $\mathcal{L}_D$ satisfy \emph{(i)}. A natural question is: which of these $\mathcal{L}_D$ must be chosen for D-FACTS deployment? To answer this question, we consider the secondary problem of cost minimization described in \emph{(ii)} .

In the absence of defense consideration, reference \cite{RogersDFACTS2008} proposed an approach to find $\mathcal{L}_D$ that minimizes the transmission losses by choosing the set of links for D-FACTS deployment that have the highest power loss to impedance sensitivity factors. We adopt a similar approach in this work. Since the primary interest of this work is the OPF cost, we compute the OPF cost to impedance (OCI) sensitivity factor for each link, i.e., $dC_{\text{OPF}}/dx_{l}.$ Then, installing a D-FACTS device on the links with the greatest absolute values of $dC_{\text{OPF}}/dx_{l}$ will achieve the least OPF cost. 

With the MTD consideration, this approach must be modified to select the set of links that have the highest OCI sensitivity values while satisfying the attack detection conditions. This can be accomplished by converting $\mathcal{G}$ to a weighted graph, where the link weights are their OCI sensitivity factors. Then, among all sets $\mathcal{L}_D$ that satisfy \emph{(i)}, we must select the set that has the greatest sum of link weights. 

Although the procedure stated above satisfies conditions \emph{(i)} and \emph{(ii)} optimally, listing all the spanning trees of $\mathcal{G}$ can be computationally complex. However, this issue can be addressed easily. Specifically, we first choose the minimum-weight spanning tree of the weighted graph, denoted by $\mathcal{L}_{\text{minsptr}}.$ Then setting $\mathcal{L}_D = \mathcal{L} \setminus \mathcal{L}_{\text{minsptr}}$ solves both \emph{(i)} and \emph{(ii)} optimally. The D-FACTS deployment algorithm is  summarized in Algorithm~1.

\begin{algorithm}
  \small
\SetAlgoLined
\KwData{Power grid graph $\mathcal{G = (N,L})$}
\KwResult{$\mathcal{L}_D$}
Set the weight of link $l \in \mathcal{L}$ as $dC_{\text{OPF}}/dx_{l}.$ \\
Compute the minimum weight spanning tree $\mathcal{L}_{\text{minsptr}}$ of $\mathcal{G}.$ \\
Set $\mathcal{L}_D = \mathcal{L} \setminus  \mathcal{L}_{\text{minsptr}}.$
\caption{\small D-FACTS Deployment Set}
\end{algorithm}
\paragraph*{Complexity of Algorithm 1}
The D-FACTS placement algorithm requires solving a Minimum Spanning Tree (MST) problem, which has a complexity of $\mathcal{O} (L \log N) $ (e.g., by using Kruskal's algorithm) \cite{BondyGraph1976}.

To conclude, consider the D-FACTS deployment set $\mathcal{L}_D$ chosen according to Algorithm~1. Assume that the defender perturbs the reactances of the set of links $\mathcal{L}_{D_w} \subseteq \mathcal{L}_D.$ Then, we have the following:
\begin{itemize}[leftmargin=*,parsep=0cm,itemsep=0cm, topsep=0cm]
\item A physical attack against a link $l $ can be detected by the BDD if the links in  $\mathcal{L}_{D_w}$ ensure that the conditions listed in Problem 1 are satisfied for that link. We will henceforth refer to such a link to being ``protected" under the MTD link perturbation set  $\mathcal{L}_{D_w}.$
\item Naturally, based on the arguments stated in this section, if $\mathcal{L}_{D_w} = \mathcal{L}_D,$ then all the links $l \in \mathcal{L}$ are protected from the physical attacks.  
\end{itemize}

We now provide extensions of the D-FACTS placement solution under some generalized conditions.

\subsection{Multiple Line Disconnections}
In the D-FACTS placement algorithm presented above, we only considered single line disconnections. However, if the attacker is able to remotely access multiple line circuit breakers, the s/he can disconnect multiple transmission lines at once. The MTD solution must be able to detect such multiple line disconnections.

The proposed D-FACTS placement algorithm remains effective in this generalized setting. Note that if the attacker disconnects multiple links, then s/he must obtain the knowledge of susceptances of all tripped branches, and the phase angles of all buses connecting to the tripped branches after the physical attack to launch an undetectable CCPA \cite{DengCCPA2017}. This in turn would require the attacker to obtain the knowledge of the line reactances of multiple loops within the power grid. 
Since our placement algorithm is designed to ensure that every loop in the graph contains at least one link with a D-FACTS device installed, the proposed algorithm can successfully invalidate the attacker's knowledge under multiple line disconnections. 

\subsection{Partial Placement of Measurement Sensors and Attacker's Access}
The system model considered thus far assumes full placement of power flow/injection measurements. However, in practice, the sensors may only be deployed partially. Moreover, the attacker may have access to only a subset of the deployed sensors, either due to his/her limited resources or due to the protection measures deployed at some of the sensors. The sensor placement and the attacker's accessed measurements affects the D-FACTS deployment solution, since they determine the knowledge of branch power flows that the attacker can obtain (which is essential for the attacker to craft an undetectable CCPA as argued in Section~\ref{sec:CCPA}). In this subsection, we investigate how the D-FACTS deployment solution is affected by these two factors.

We assume that the system operator deploys the power flow/injection sensors to satisfy the basic observability condition, i.e., to ensure that the observability matrix has full rank \cite{AburExposito2004}. Under this assumption, if the attacker has access to all the deployed sensors, then the proposed D-FACTS deployment in Algorithm~1 remains valid and optimal. Indeed, in this scenario, the attacker can derive the power flows on all branches of the power grid (fully observable system). Thus, the D-FACTS deployment solution will not change.

However, if the attacker can access only a subset of the deployed sensors, then s/he cannot derive the power flows on all the branches. Thus, the attacker will not be able to use the paths that contain the \emph{unobservable} branches to derive the phase angle difference between the disconnected nodes. In this scenario, the number of D-FACTS devices that need to be deployed (for MTD) can be significantly reduced. The system operator can first determine the unobservable branches and the \emph{observable islands} \cite{AburExposito2004} from the attacker's perspective based on the measurement sensors that s/he can access (e.g., knowledge of unprotected sensors). Note that observable islands in a power system can be determined in an efficient manner using existing numerical methods (see e.g., \cite{Castillo2006}). Let us denote these observable islands by $\mathcal{G}_i = \{ \mathcal{N}_i, \mathcal{L}_i \} , i = 1,\dots,K,$ where $K$ denotes the number of islands. 

Then, the D-FACTS deployment set can be determined by ensuring that there are no loops within any observable island $\mathcal{G}_i, i = 1,\dots,K$. The rationale is that, if the attacker disconnects a link between any buses that belong to different observable islands, then every alternative path between the disconnected nodes will necessarily involve an unobservable link. Therefore, s/he cannot launch an undetectable CCPA against such links. In other words, the attacker can launch undetectable CCPAs only against the links that connected two nodes within the same observable islands. Therefore, to defend against these CCPAs, it suffices to ensure that each observable island is loopless. The overall solution is summarized in Algorithm~2.


\begin{algorithm}
	\small
	\SetAlgoLined
	\KwData{Power grid graph $\mathcal{G = (N,L})$, Attacker's sensor access set $\mathcal{P}_A$}
	\KwResult{$\mathcal{L}_D$}
	Set the weight of link $l \in \mathcal{L}$ as $dC_{\text{OPF}}/dx_{l}.$ \\
	Using $\mathcal{P}_A,$ compute the ubobservable branches and enlist the set of observable islands $\mathcal{G}_i = \{ \mathcal{N}_i, \mathcal{L}_i \} , i = 1,\dots,K$. \\
	For each observable island, compute the minimum weight spanning tree $\mathcal{L}_{i,\text{minsptr}}$ of the corresponding graph. \\
	Set $\mathcal{L}_D = \cup^K_{i = 1}\mathcal{L}_i \setminus  \mathcal{L}_{i,\text{minsptr}}.$
	\caption{\small D-FACTS Deployment Set With Partial Sensor Placement}
\end{algorithm}

\paragraph*{Complexity of Algorithm 2}
If the attacker has access to $M$ measurements (where $M \leq 2L+N$, then the complexity of computing the observable islands using numerical methods is $\mathcal{O} (MN)$ \cite{Castillo2006}. The overall complexity of finding MSTs in all the observable islands is $\mathcal{O} (N L).$

Note that in the general case where the there are no protected sensors or if the system operator has no knowledge of the sensors that can be accessed by the attacker, our original MTD deployment (Algorithm~1) remains valid. An illustration of D-FACTS devices under partial sensor deployment is presented in Section~\ref{sec:sims}.

\subsection{Attack Detection Under the AC Power Flow Model}
The proposed MTD design will also be effective under the AC power flow model. This can be explained as follows. Recall that in general, an undetectable CCPA can be constructed as $\av = \zv - \zv_p,$ where $\av$ is the attack vector to be injected, $\zv$ are the original measurements and $\zv_p$ are the measurements following the physical attack  (line disconnection). In the case of the DC power flow model, as shown in \cite{DengCCPA2017}, the attack vector $\av$ can be computed in an optimized manner using the  knowledge of the reactances of only a few links with in the grid (as explained in Section III of the paper). In contrast, no optimal methods are known to compute the attack vector $\av$ for the AC power flow model. Thus, the attacker will have first to recompute the AC power flow equations to obtain $\zv_p$, and subsequently obtain the attack vector $\av.$ This in turn would require the knowledge of the reactances of all the links in the power grid. Thus, our proposed MTD design will remain effective because it invalidates the attacker's knowledge of the branch reactances of the links within $\mathcal{L}_D.$  

Alternately, the recent work \cite{ChungLocal2019} shows that the attacker can use line outage distribution factors (LODFs) to compute $\zv_p$. However, the computation of LODFs also require the knowledge of the line reactances of all the links in the grid \cite{GuoLODF2009} (note that \cite{ChungLocal2019} also assumes that the attacker has knowledge about the topology of the entire system). Our extensive simulation results presented in Section~\ref{sec:sims} verify that the proposed MTD design remains effective under the AC power flow model under both full power flow recomputation as well as using LODFs.


\section{Game Theoretic MTD Robust Strategies }
\label{sec:Game}
MTD perturbations incur an operational cost, and perturbing the reactances of a large set of links may not be cost effective. Instead, we propose to protect only a subset of links from physical attacks depending on the operational system state, as well as the perceived threat to those links. This is approached using a game-theoretic formulation presented next. 

\subsubsection{Zero-sum Game Formulation}
We define the strategic interactions between the attacker and the defender as a two-player {zero-sum} game. To formalize this, we define the game as a triplet $\Gamma \triangleq \left( \{D,A\}, \{\mathcal{S}_D, \mathcal{S}_A\}, \{u_D, u_A\} \right)$
in which the components are: (i) the set of players $\{D,A\}$;  (ii) $\mathcal{S}_D$ and $\mathcal{S}_A,$ the sets of actions that  defender and attacker can take respectively; and  (iii) the payoffs of the players $u_k: \mathcal{S}_D \times \mathcal{S_A} \rightarrow \mathbb{R}$ for $k\in \{D,A\},$ where $u_k (s_D, s_A)$ measures the benefit obtained by player $k$ when the action profile that has been played is $s=(s_D,s_A)$. In a zero-sum game, the payoffs are opposite and $u_A(s_D,s_A) = - u_D(s_D,s_A)$ such that a cost for the defender is a benefit for the attacker and vice-versa.


We denote the attacker's and the defender's action sets by $\mathcal{S}_A  = \{ a_0, a_1, \dots,  a_{N_A-1} \}$ and $\mathcal{S}_D  = \{ d_0, d_1, \dots,  d_{N_D-1} \}$ respectively, where $N_A$ and $N_D$ are the cardinality of the sets $\mathcal{S}_A$ and $\mathcal{S}_D$ respectively. The attacker's action set is the subset of links it disconnects physically. We denote the set of links disconnected by the attacker under action $a_i$ by $\mathcal{L}_{a_i},$ where, $\mathcal{L}_{a_i} \subseteq \mathcal{L}, \ i = 0,1, \dots,N_{A}-1.$ 
The action $a_0$ corresponds to the case when the attacker does not attack any link. Note that the attacker's action set $\mathcal{S}_A$ may also include multiple line disconnections. The defender's action is to select a subset of links within $\mathcal{L}_D$ whose reactances will be perturbed. We denote the set of links chosen by the defender under action $d_i$ by  $\mathcal{L}_{d_i},$ where, $\mathcal{L}_{d_i} \subseteq \mathcal{L}_D, \ i = 1, \dots,N_{D}-1.$
The action $d_0$ corresponds to the case when the defender does not perturb the reactance of any link.



Next, we characterize the payoff $u_d(s_D,s_A)$. If the attacker's chosen action $s_A$ includes a link $l$ that is protected by the defender (via MTD), then the CCPA will be detected by the BDD, and the operator can quickly restore the link to avoid any further damage. For instance, the defender can quickly restore the circuit breaker of the disconnected link to a closed position. On the other hand, if the attacker disconnects a subset of links that are not protected, then the CCPA will go undetected. The link disconnection will result in redistribution of power flows. Consequently, the power flow on some of the links may exceed the corresponding thermal limits. The system operator will notice the power flow violations and initiate a generator dispatch/load shedding to resolve this issue, which in turn rebalances the power flows and rectifies the overflows. We denote the cost of load shedding at bus $i$ by $C_{i,s} (L^s_{i}),$ where $L^s_{i} (\leq L_i)$ is the quantity of load that is shed, and $\dv_s = [D^s_{1},\dots,D^s_{N}].$


Let $C_{\text{OPF}} (a_m,d_n)$ denote the OPF cost when the attacker takes an action $a_m$ and the defender takes an action $d_n$, which can be computed as follows:
\beqa
 C_{\text{OPF}}({a_m,d_n})   =  & \dsp \min_{ \gv, \dv_s} &  \sum_{i \in \mathcal{N}} C_{i,g} (G_{i}) + C_{i,s} (D^s_{i}) \label{eqn:OPF_shed}
    \\ 
& s.t. &  \gv^P_{a_m,d_n}(\pv_G,\thetav,\vv,\xv) = 0, \nonumber
 \\
 & s.t. &  \gv^Q_{a_m,d_n}(\pv_G,\thetav,\vv,\xv) = 0, \nonumber
 \\
 & & | \gv - \gv^{\prime} | \leq \Delta^{\max}_g \nonumber \\
& & \fv \in \mathcal{F}, \gv \in \mathcal{G}, \nonumber
\eeqa 
where $\Bm_{a_m,d_n}$ is given by $\Bm_{a_m,d_n} =  \Am_{a_m,d_n} \Dm_{a_m,d_n} \Am_{a_m,d_n}^T.$ Here, $\Am_{a_m,d_n}$ is the bus-branch connectivity matrix when the attacker and the defender choose actions $a_m$ and $d_n$ respectively. These quantities are computed as in Algorithm~2.

\begin{algorithm}
  \small
\SetAlgoLined
\KwData{$a_m,d_n$}
\KwResult{$C_{\text{OPF}}({a_m,d_n})$}
Set branch reactances to $\xv_{d_n}.$  \\
Set $\Am_{a_m,d_n} = \Am_{a_0,d_0}.$ \\
Solve \eqref{eqn:OPF_shed} to obtain $C_{\text{OPF}}({a_0,d_n})$.
\\
\eIf{attack is successful}{
Set $\Am_{a_m,d_n}$, $\Dm_{a_m,d_n}$ by removing the branches  $\mathcal{L}_{a_m}.$
Solve \eqref{eqn:OPF_shed} to compute $C_{\text{OPF}}({a_m,d_n}).$} {Set $C_{\text{OPF}}({a_m,d_n}) = C_{\text{OPF}}(a_0,d_n).$} 
\caption{\small Payoff Computation}
\end{algorithm}
\paragraph*{Complexity of Algorithm 3}
The payoff computation algorithm requires solving $N_A N_D$ DC OPF problems. Note that DC OPF problems can be solved efficiently, e.g., by interior point methods in polynomial time.

Based on the above, the defender's payoff is given by
\begin{equation*}
u_D(s_D,s_A) = \left\{ 
\begin{array}{ll}
C_{\text{OPF}}(d_0, a_0) - C_{\text{OPF}}(s_D, a_0),   & \text{if} \ \mathcal{I}_S = 0  \\
C_{\text{OPF}}(d_0, a_0) - C_{\text{OPF}}(s_D, s_A),  & \text{if} \ \mathcal{I}_S = 1.
\end{array}\right.
\end{equation*}
The term $\mathcal{I}_S$ is an indicator variable  to represent the success ($\mathcal{I}_S = 1$) or failure of an attack  ($\mathcal{I}_S = 0$).
Both players aim to choose their actions such that their own payoff is maximized 
and we can see that the two players have contradictory objectives. 

{The above payoffs can be explained by looking at them as negative costs.} First, $C_{\text{OPF}}(d_0,a_0)$ denotes the benchmark operating cost of the defender when none of the players takes an action to either disrupt or defend the system. The term $C_{\text{OPF}}(s_D, s_A) - C_{\text{OPF}}(d_0, a_0)$ denotes the the additional cost incurred by the defender and caused by a successful attack, when the attacker chooses $s_A$ and the defender chooses $s_D$.
The term $C_{\text{OPF}}(s_D, a_0) - C_{\text{OPF}}(d_0, a_0)$ represents the additional cost incurred by the defender for choosing an action $s_D$ against an unsuccessful attack $s_A$. {Hence, the aim of the defender is to minimize these costs, whereas the attacker will seek to maximize them.}

\subsubsection{Nash Equilibrium Solution}
In such an interactive situation, the natural solution is the Nash equilibrium (NE). However, the game $\Gamma$ above is discrete and finite and may not have a NE solution in pure strategies. 
Instead, it always has at least one mixed-strategy NE  \cite{Tirole1991}, which is the NE of its extension to mixed strategies. The latter is defined as follows: $\widetilde{\Gamma} \triangleq \left( \{D,A\}, \{ \Delta_D, \Delta_A\}, \{\tilde{u}_D, \tilde{u}_A\} \right)$.
The action sets of the extended game $\widetilde{\Gamma}$ are the probability simplices of dimension $N_k$, $k\in\{D,A\}$:
$\Delta_k = \left\{ p_k \in \mathbb{R}^{N_k}_+ \left| \sum_{j=0}^{N_k} p_{k,j} =1 \right. \right\}$
where $p_k =(p_{k,0}, \hdots, p_{k,N_k-1})$ is the discrete probability vector of player $k$ such that $p_{D,j}$ and $p_{A,j}$ represent the probability of choosing the action $d_j$ by the defender and the probability of choosing the action $a_j$ by the attacker, respectively. The modified payoffs are simply the resulting expected payoffs following the randomization of play: 
\begin{equation}
\tilde{u}_k (p_D,p_A) = \sum_{j=0}^{N_D-1} \sum_{i=0}^{N_A-1} u_k(d_j, a_i) \ p_{D,j} \ p_{A,i}.
\end{equation}

The mixed NE is a stable state to unilateral deviations, which means that no player can benefit by deviating from their NE strategy individually.
\begin{definition}
A strategy profile $(p_D^*, p_A^*)$ is a mixed NE for the game $\Gamma$, iff the following conditions are met:
$\tilde{u}_D(p_D^*, p_A^*)  \geq  \tilde{u}_D(p_D, p_A^*), \ \forall \ p_D \in \Delta_D,$ and 
$\tilde{u}_A(p_D^*, p_A^*)  \geq  \tilde{u}_A(p_D^*,p_A), \ \forall \ p_A \in \Delta_A.$
\end{definition}

The mixed NE can also be characterized by the Von-Neumann indifference principle \cite{Tirole1991}, which requires that: i) player $k$ is rendered indifferent between its pure actions played at the NE (with strictly positive probability), by the choice of the other player $p_{-k}$, for all $k\in \{D,A\}$; and ii) the actions that are not played at the NE give strictly lower payoffs than the ones that are played, for both players. 

This characterisation allows one to compute the mixed NE in a straightforward manner (by solving a linear system of equations), if the exact face of the simplex $\Delta_D \times \Delta_A$ on which the NE $(p_D^*, p_A^*)$ lies is known. Computing this among $2^{N_D+N_A}$ possible faces is a well-known difficult problem: the Lemke-Howson algorithm is the best combinatorial algorithm \cite{roughgarden-2010} and is PSPACE-complete.\footnote{{PSPACE-complete problems are the hardest problems in polynomial space (PSPACE, i.e.,  problems that can be solved using an amount of memory that is polynomial in the input length) and are such that every other PSPACE problem can be transformed to it in polynomial time. They are suspected to lie outside of the set of NP-hard problems but this remains to be proven.}}. 
In the worst case scenario, it performs as poorly as exhaustive search (complexity grows exponential with the dimension of the action profile set), but in general it can be quite efficient.

Aside from complexity, the major drawback of the Lemke-Howson algorithm is that it requires \emph{perfect and complete information} of the game $\mathcal{G}$ at both players. In an adversarial setting, assuming that the players know precisely the action sets of their opponent and the payoff function is not realistic. Instead, we suggest to exploit iterative learning processes that allow the players to compute their NE strategies by learning from their past interactions.

\subsubsection{Machine Learning to Solve the Game}

{
We focus here on the \emph{exponential weights} for exploration and exploitation (EXP3) algorithm, also known as the multiplicative weights, which is a ubiquitous iterative decision process that has been repeatedly discovered in machine learning, optimization and game theory \cite{auer-2003, hazan-2012}; having wide applications such as: AdaBoost for classification and prediction \cite{hastie-2009}, pooling problems for blending industries \cite{mencarelli-2017}, graphical model learning \cite{klivans-2017}, and learning the NE in certain types of games \cite{auer-2003, mertikopoulos-2019} (e.g., potential games, two-player zero-sum games, etc.).

The main idea of EXP3 is that, at each iteration $t$, the decision agent $k$ chooses a random action $s_{k,t}$ following the probability distribution $p_{k,t} = [p_{k,t}(s)]_{s\in \mathcal{S}_k}$. As a result, the agent observes the value of its payoff: $u_k(s_{k,t}, s_{-k,t})$, based on which the cumulative scores of all actions are updated. Since only the payoff of the realized action can be observed, we have to estimate the payoffs of the unplayed ones. For this, the following pseudo-estimator can be built, for any $s \in \mathcal{S}_k$ 
\begin{equation}
\label{eq:est_pay_exp3}
\hat{u}_{k,t}(s) = \frac{u_k (s_{k,t}, s_{-k,t} ) \mathbb{I}\{s=s_{k,t}\}+ \beta_t}{p_{k,t}(s)}
\end{equation}
where the first term,  $u_k (s, s_{-k,t} ) \mathbb{I}\{s=s_{k,t}\} /p_{k,t}(s)$ is known as \emph{importance sampling} and represents an unbiased estimator of $u_k(s, s_{-k,t})$ for any $s$ at time $t$. To control the variance of this estimator, the bias $\beta_t >0$ is introduced \cite{auer-2003}.

The cumulative score of action $s$ is defined as: $G_{k,t}(s) = \sum_{\tau=1}^{t} \eta_{\tau} \hat{u}_{k,\tau} (s)$, which measures how well the explored actions have performed in the past.
Then, these cumulative scores are mapped on the probability simplex via a well chosen exponential map
\begin{equation}
\label{eq:probas_exp3}
p_{k,t+1}(s) =\gamma_t \frac{1}{|\mathcal{S}_k |}  + (1-\gamma_t) \frac{\exp( G_{k,t}(s))}{ \sum_{r\in \mathcal{S}_k} \exp(G_{k,t} (r))}, \ \ \forall s,
\end{equation}
where $\gamma_t \in (0,1]$ and $\eta_t>0$ are learning parameters and $|\mathcal{S}_k |$ denotes the cardinal of the set $\mathcal{S}_k$. 
Intuitively, this means that the actions that have been performed well in the past are played with relatively higher probability, without discarding completely poor or unexplored actions that may perform well in the future; this illustrates the \emph{data exploration vs. exploitation tradeoff}. The updated probability distribution $p_{k,t+1}$ will hence be used by player $k$ to generate the random action at the next iteration $t+1$ and so on. The details are provided in Algorithm~3 below. 

\begin{algorithm}
  \small
\SetAlgoLined
\KwData{$\gamma_t, \beta_t, \eta_t, u_k(s_{k,t}, s_{-k,t})$}
\KwResult{$\overline{p}_{k,t}$}
Initialize $t=1$, $p_{k,1}(s) =\frac{1}{|\mathcal{S}_k|}$, $G_{k,0}(s) = 0$ for all $s\in \mathcal{S}_k$\\
\While{$\overline{p}_{k,t}$ has not converged}{
Draw random action $s_{k,t}$ following distribution $p_{k,t}$\\
Observe payoff $u_k (s_{k,t}, s_{k,-t})$\\
Compute pseudo-estimations $\hat{u}_{k,t}(s), \ \forall s\in\mathcal{S}_k$ as in eq. \eqref{eq:est_pay_exp3}\\
Update cumulative scores:
$$G_{k,t}(s) = G_{k,t-1}(s) + \eta_t \ \hat{u}_{k,t}(s), \ \forall s\in\mathcal{S}_k$$\\
Update probability distribution $p_{k, t+1}$ as in eq. \eqref{eq:probas_exp3}\\
Compute empirical frequency $\overline{p}_{k,t}$ as in \eqref{eq:empirical_exp3}\\
Step $t\leftarrow t+1$
}
\caption{\small EXP3 to compute the NE at player $k$}
\end{algorithm}

Notice that the three parameters of the algorithm: $\beta_t$, $\gamma_t$, and $\eta_t$ have to be very carefully tuned. Indeed $\beta_t>0$ controls the bias vs. variance tradeoff of the payoff estimator $\hat{u}_{k,t}(s), \forall s$ necessary because of the limited information available to player $k$. Indeed, only the value of the payoff $u_k(s_{k,t}, s_{-k,t})$ is observed while the opponent's action $s_{-k,t}$ is unknown. 

The parameters $\gamma_t\in (0,1]$ and $\eta_t>0$ impact the exploration vs. exploitation tradeoff. For $\gamma_t=1$ (or $\eta \ll1$) there is no data exploitation but pure exploration (no learning from past results), as the actions are drawn following the uniform distribution; whereas for $\gamma_t \ll  1$, the exploration term is reduced in favour of the exponential weights. For $\eta_t \gg 1$, the algorithm stops exploring at the first chosen action.

In \cite{auer-2003}, it was shown that the EXP3 algorithm above converges to the mixed Nash equilibria. The theoretical result is reported below as in \cite{lazaric-2017}, for the sake of simplicity.
\begin{proposition}
If each player of the zero-sum game $\Gamma$ runs the EXP3 algorithm with parameters:\\ $\gamma_t=\sqrt{|\mathcal{S}_k| \log |\mathcal{S}_k| / t}$, $\eta_t \approx \beta_t= \sqrt{2 \log |\mathcal{S}_k|/ (t |\mathcal{S}_k|) }$, then their empirical frequencies 
\begin{equation}
\label{eq:empirical_exp3}
\overline{p}_{k, t} = \frac{1}{\sum_{\tau=1}^{t} \eta_{\tau} }\sum_{\tau=1}^{t} \eta_{\tau} p_{k,\tau}
\end{equation}
converge asymptotically to the set of the mixed NEs. The convergence rate is $\mathcal{O}(T^{-1/2})$, where $T$ is the horizon of play.
\end{proposition}

\paragraph*{Information requirement of EXP3}
The major advantage of the EXP3 algorithm is the fact that it does not require the players to have complete and perfect knowledge of the game (as opposed to the Lemke-Howson algorithm), 
neither the actions chosen by their adversary. Both players only observe the payoff value as a result of their chosen actions at each iteration. Moreover, unlike the Lemke-Howson algorithm, EXP3 is shown to be robust to estimation errors of the payoff observation \cite{mertikopoulos-2019}: under mild assumptions on the error process (zero-mean and finite variance) EXP3 retains its $\mathcal{O}({T}^{-1/2})$ convergence speed even when only a noisy observation of the payoff is available at each iteration.

\paragraph*{Complexity of EXP3}
One iteration of EXP3 has a relatively low complexity, increasing only linearly with the number of possible actions, i.e., $\mathcal{O} ( |\mathcal{S}_k| )$ for player $k$. Also, the convergence speed of EXP3 is $\mathcal{O}(T^{-1/2})$, which implies that to reach an $\varepsilon$-neighborhood of the Nash equilibrium solution, the algorithm requires $1/\varepsilon^2$ iterations. We note that in the adversarial setting, this convergence speed is optimal and cannot be improved because of the limited available information \cite{auer-2003}.

At this point, some comments on the convergence time of EXP3 algorithm are in order. First, the convergence time depends on the defender's action set $\mathcal{S}_D$ and also on the attacker's action set $\mathcal{S}_A$, which can be exponentially large when considering multiple line disconnections. Second, the convergence time $\mathcal{O}(T^{-1/2})$ may seem long. Nevertheless, for the proposed MTD application, these factors will not be a major issue, for two reasons: (i) As we have explained in Section~\ref{sec:MTD}-A, the time period between successive MTD perturbations can be reasonably long (e.g., in \cite{LakshDSN2018}, we show that hourly MTD perturbations might be realistic for practical systems). (ii) For several combinations of multiple line disconnections, the net measurements following the CCPA, i.e., $\zv_p + \Delta H \theta_p$ turns out to be significantly different from the original measurements $\zv$ (before the CCPA). Such attacks can be easily detected by the system operator, by monitoring the change in the measurements over successive monitoring periods. Note that under normal system operation, the measurements change only due to the fluctuation in the system load, which tends to be smooth. Thus, any abrupt change in the system measurements can trigger an alarm. Thus, many of the multiple line disconnections need not be considered in the NE computation. We verify this using simulations presented in Section~\ref{sec:sims}, Fig.~\ref{fig:Diff_Mes} .

At last, if one were to investigate AC OPF considerations, this would only change the computation of $C_{\mathrm{OPF}}(a_m,d_n)$ in equation (4). Algorithm 3 would be impacted in the sense that it would require solving $N_A N_D$ AC OPF problems as opposed to $N_A N_D$ DC OPF problems. Although this would imply a change in the values of the payoffs $u_D (s_D, s_A)$ and $u_A(s_D,s_A)$, this would not change the structure of the game $\Gamma$ under study, which remains a \emph{two player zero-sum game with discrete and finite sets of actions}. Once the new values of the payoffs are computed, the entire analysis of the Nash equilibrium solution provided above still stands. Furthermore, the machine learning approach to find the Nash equilibrium also stands.

\vspace{-0.1in}
\section{Simulation Results}
\label{sec:sims}
Below, we present our simulation results, performed with the MATPOWER toolbox, to show the effectiveness of the proposed defense. 

\paragraph{MTD Attack Detection}
First, we examine the MTD's capability in detecting CCPAs using IEEE 14-, 39- and 118-bus systems. As proposed in Section~\ref{sec:MTD_Soln}, we solve the minimum weight feedback edge set problem for the graph corresponding to the IEEE bus system and determine $\mathcal{L}_D$. Since there is no closed form expression for the weights $dC_{\text{OPF}}/dx_l,$ we compute them by simulations. We then perturb the reactances of all the links in the set  $\mathcal{L}_D.$ 
First, we simulate physical attacks against a single randomly chosen link in the bus system and inject a corresponding CCPA of the form $\av = \Delta \Hm \thetav_p,$ where both $\Delta \Hm$ and $\thetav_p$ are computed using outdated knowledge of the system. Following the arguments presented in Section \ref{sec:MTD_Soln}, we note that any random value of reactance perturbation will ensure that the CCPAs no longer remain undetectable (i.e., its detection rate will be greater than the false positive rate, since the BDD residual will be necessarily non-zero). However, in addition, the system operator may also want to ensure that the CCPAs will also detected with a high probability (i.e., detection rate close to $1$). We plot the BDD's attack detection probability for each case  in Fig.~\ref{fig:Result1} as a function of the percentage change in line reactances. It can be observed that that $10-20 \%$ perturbation in the line reactances will ensure that CCPAs are detected with a very high probability for most IEEE bus systems.

Next, we verify the effectiveness of the proposed approach in detecting CCPAs against multiple link disconnections. We disconnect links $\{2,5 \}$ and $\{2,5,6 \}$ in the IEEE-14 bus system are inject the corresponding cyber attacks (constructed based on the outdated reactance knowledge) to mask the effect of the physical attack. The results are plotted in Figure~\ref{fig:Multiple_Links}. We observe that the proposed MTD can effectively detect CCPAs in this case as well.

We also compare the attack detection rate achieved by the proposed D-FACTS placement strategy against two other placement strategies: (i) D-FACTS placement that minimizes the OPF cost only (i.e., obtained by choosing only the links with highest $dC_{\text{OPF}}/dx_l,$ values ) (ii) random D-FACTS placement (in which the D-FACT links are chosen at random). We perform simulations using the IEEE-14 bus system. For optimal D-FACTS placement (i.e., obtained by Algorithm~1), D-FACTS devices must be placed on $7$ links given by $1,3,5,8,9,18,19.$ For fair comparison, we also place $7$ D-FACTS devices for the minimum cost and random D-FACTS placement. The detection rate achieved by the three placement strategies is plotted in Fig.~\ref{fig:Compare_placement} for different CCPAs against different links of the grid. It can be observed that the optimal D-FACTS placement effectively detects all physical attacks, where as the other two strategies are only effective against a few link disconnections only.  

\begin{figure}[!t]
\centering
\includegraphics[width=0.45\textwidth]{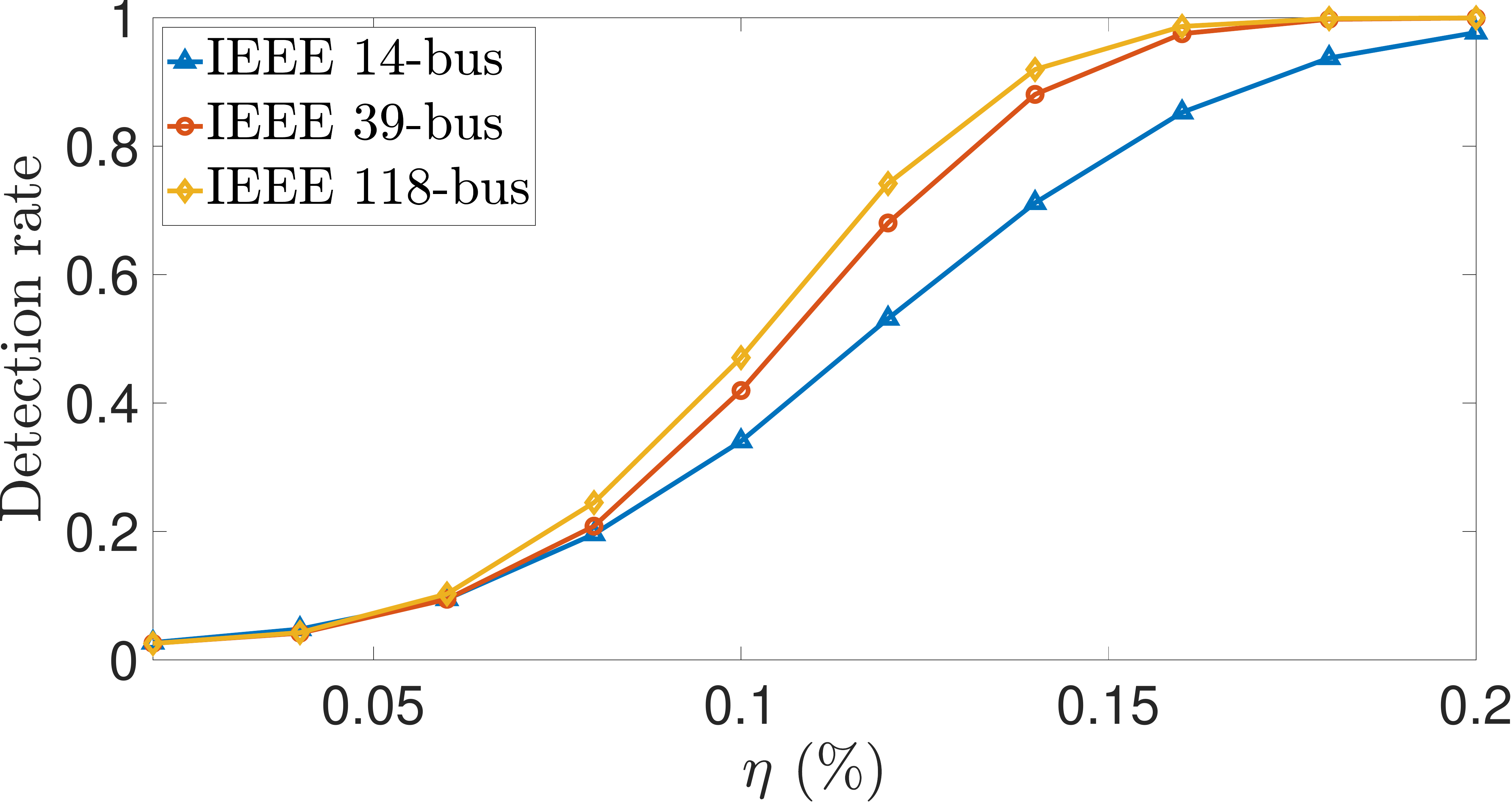}
\caption{Attack detection probability for single line disconnections as a function of the percentage change ($\eta$) in the link reactance.}
\label{fig:Result1}
\vspace{-0.2 cm}
\end{figure}

\begin{figure}[!t]
  	\centering
  	\includegraphics[width=0.45\textwidth]{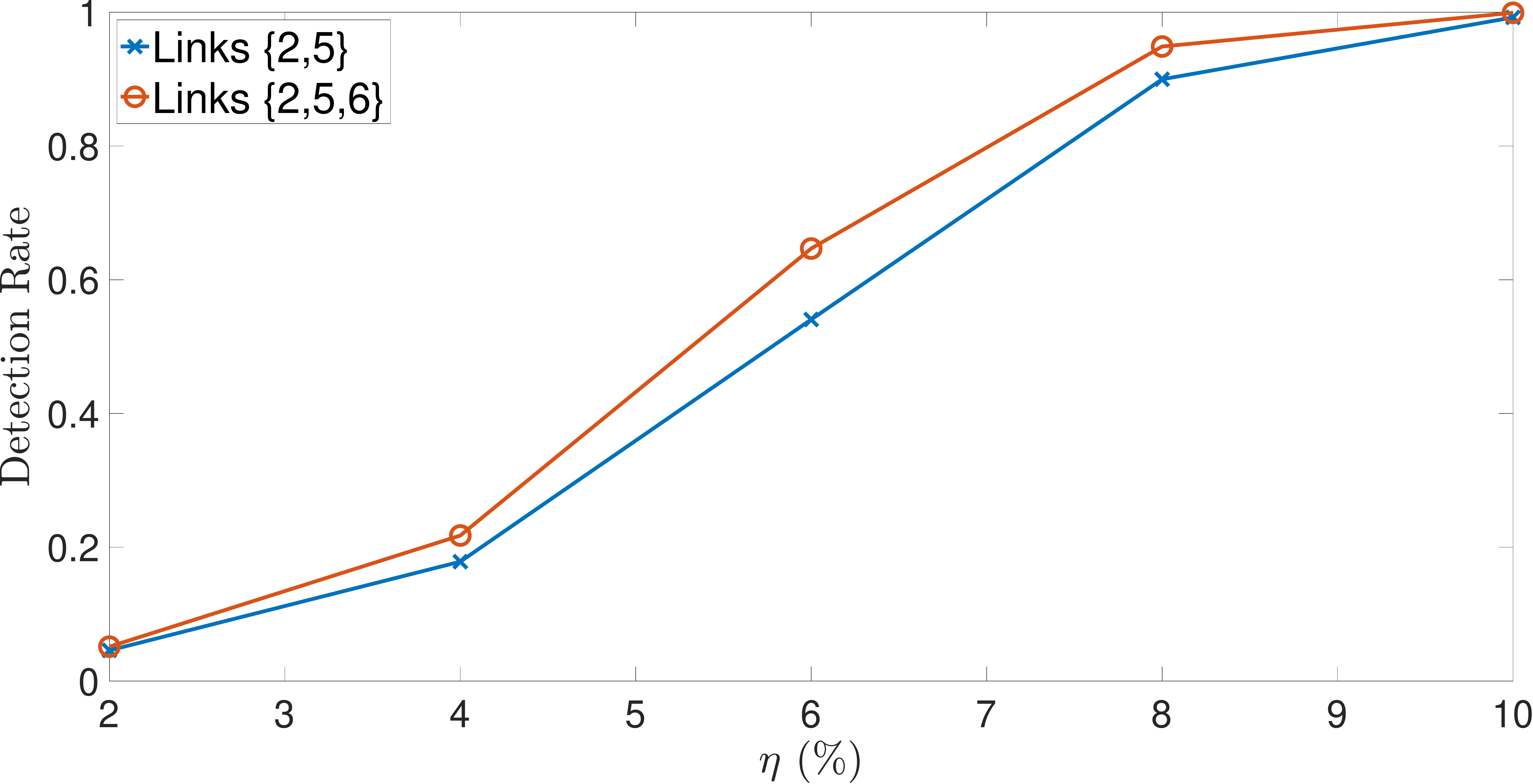}
  	\caption{Attack detection probability for multiple line disconnections in the IEEE-14 bus system.}
  	\label{fig:Multiple_Links}
  	\vspace{-0.2 cm}
  \end{figure}

\begin{figure}[!t]
	\centering
	\includegraphics[width=0.45\textwidth]{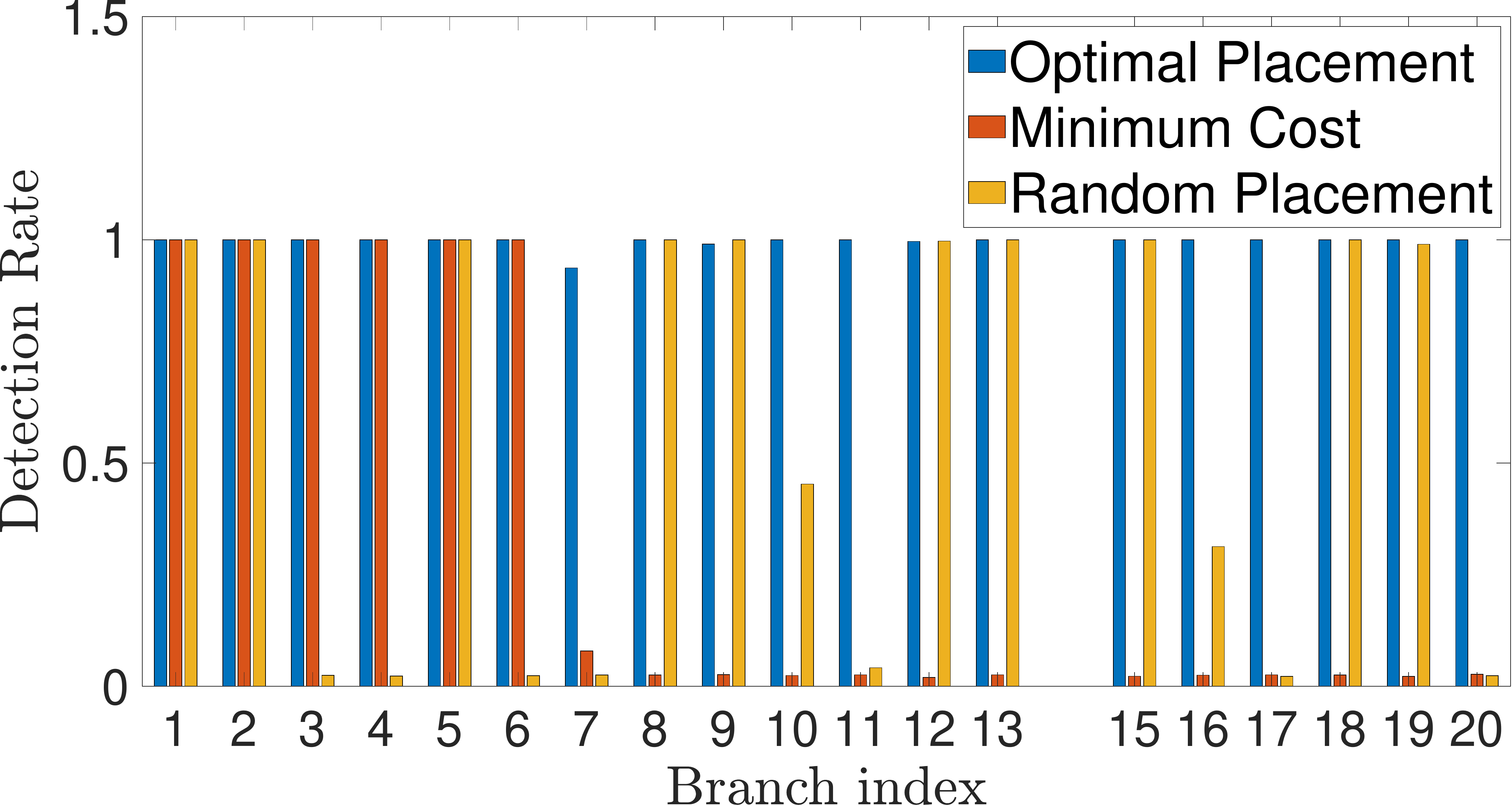}
	\caption{Attack detection probability for different D-FACTS placement strategies. The x-axis label denotes the index of the disconnected link under the CCPA.}
	\label{fig:Compare_placement}
	\vspace{-0.2 cm}
\end{figure}


We also enlist the size of the D-FACTS deployment sets (that can detect any CCPA) in Table~\ref{tbl:DFACT_Size}. The proposed approach enables the defender to protect the power grid with relatively few D-FACTS devices. We can also conclude that $|\mathcal{L}_D|$ depends on the grid's actual topology and not just on its size (e.g., $|\mathcal{L}_D| = 15$ for the 24 bus system, whereas $|\mathcal{L}_D| = 9$ for the 39-bus system).

We also investigate the D-FACTS deployment set under partial sensor placement and limited access of the attacker. We conduct simulations using the IEEE-24 bus system. The simulation settings are listed in Table~\ref{tbl:Partial}. The sensor deployment set is selected to ensure full system observability. However, the attacker has access only a subset of the system, due to which several branches become unobservable. Among all the observable islands, only the island \{1,2,4,5\} contains a loop. Deploying D-FACTS device on link $1-2$ ensures that this island is loopless. 

\begin{table*}[!t]
\begin{center}
 \begin{tabular}{||m{3cm}|m{6cm}| m{3cm}| m{3cm}||} 
 \hline  
\center{  Measurements } & \center{Index}  & \center{  $\mathcal{L}_D$ (Full access) } &  $\mathcal{L}_D$ (Partial access) \\[2 ex]  
 \hline\hline 
\center{  Power flow } &  \center{ Links $ \{ 1, 3, 4, 7,13, 15, 17, 23, 32, 34, \underline{6}, \underline{18},  \underline{29}, \underline{33} $ \} } & \multirow{2}{*} \text{Links \{1,4, 6,7,12,14,18,19,22,25,31\} } & \multirow{2}{*} \text{\center{Link $\{1  \}$}}  \\ [2 ex] 
  \cline{1-2}
\center{  Power injection }  & \center{ Nodes $\{2,7,10,13,15,19,22,24, \underline{3} \} $ } &  &  \\ [2 ex]
 \hline
 \hline
\end{tabular}
\end{center}
\caption{D-FACTS placement under partial placement of sensors for the IEEE-24 bus system. Underline indicates sensors that are inaccessible to the attacker. Observable islands with attacker's accessed measurements \{1,2,4,5\}, \{3,15,24\}, \{6\}, \{7,8,9,10,12\}, \{11\}, \{13\}, \{14,16,19,20\}, \{17,21,22\}, \{18\} and \{23\}. }
\label{tbl:Partial}
\end{table*}

\begin{table}[!t]
 \begin{center}
 \begin{tabular}{||c | c | c | c ||} 
 \hline
 Bus system & $|\mathcal{L}|$ & $|\mathcal{L}_D|$  \\ [0.5ex] 
 \hline\hline
 IEEE 9-bus system & 9 & 1  \\ 
 \hline
 IEEE 14-bus system & 20 & 7  \\
 \hline
 IEEE 24-bus system & 38 & 15  \\
 \hline
 IEEE 39-bus system & 36 & 8  \\
 \hline
 IEEE 118-bus system & 186 & 62  \\
  [1ex] 
 \hline
\end{tabular}
\end{center}
\caption{Size of the D-FACTS deployment set $|\mathcal{L}_D|$.}
\label{tbl:DFACT_Size}
\vspace{-0.2 cm}
\end{table}


\begin{figure}[!t]
\centering
\includegraphics[width=0.45\textwidth]{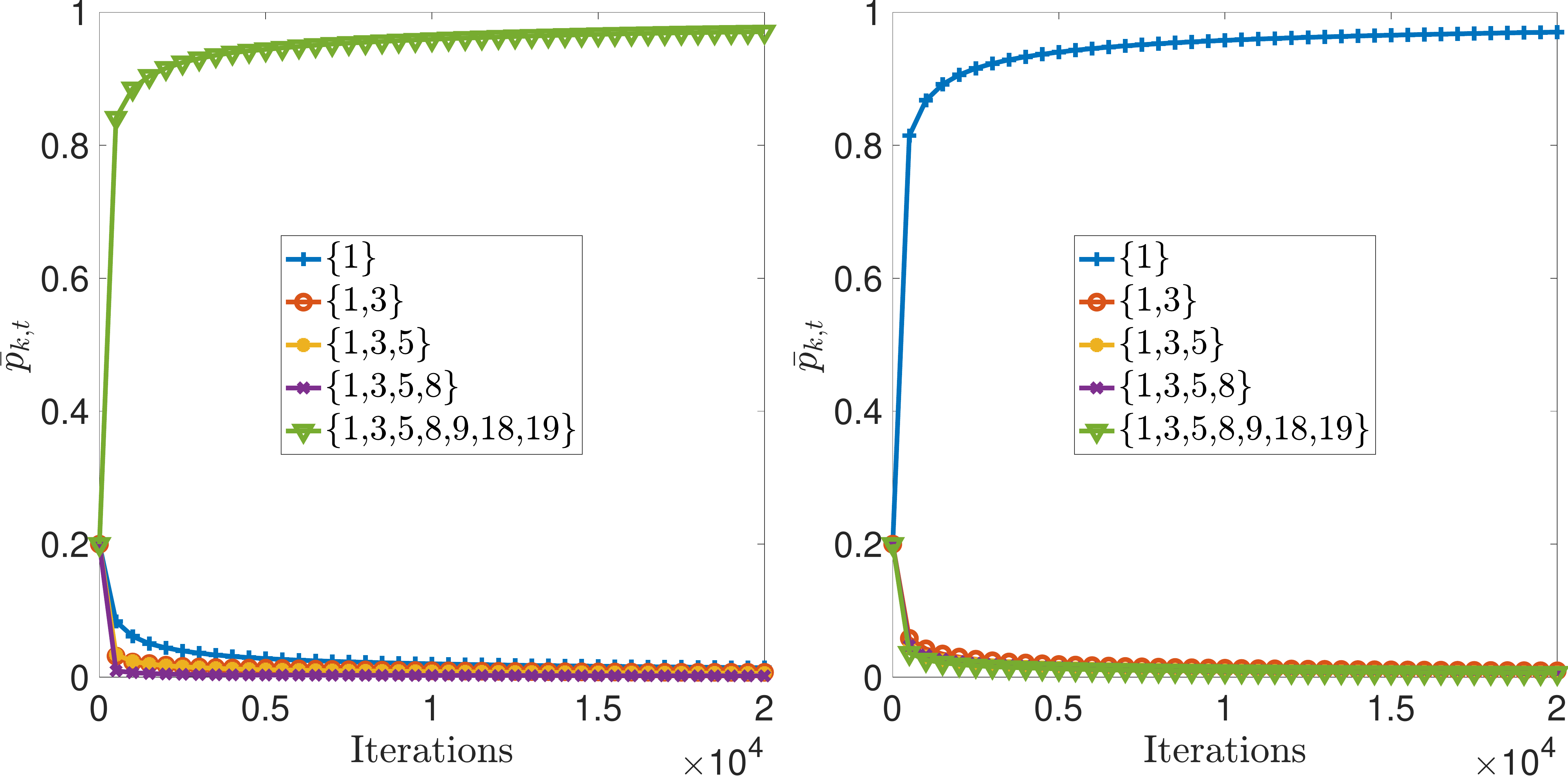}
\caption{Convergence of $\bar{p}_{k,t}$ to the NE for IEEE-14 bus system. Left: Heavily loaded system. Right: Lightly loaded system.}
\label{fig:Exp_conv}
\end{figure}

\begin{table}[!t]
 \begin{center}
 \begin{tabular}{||c | p{2cm} | p{1 cm} | p{1cm}  |c ||} 
 \hline
 Load scenario & NE D-FACTS perturbation set (Lemke-Howson) & NE defense cost &  Full defense cost \\ [0.5ex] 
 \hline\hline
 Scenario~1 & \{1,3,5,8,9,18,19\} & 9.85 \% &  9.85 \%   \\ 
 \hline
 Scenario~2 &  \{1 \} & 0.8 \%  & 4.37  \%  \\
 \hline
\end{tabular}
\end{center}
\caption{D-FACTS perturbation set and defense cost (the \% increase in OPF cost) at the NE for different system loads.}
\label{tbl:NE}
\vspace{-0.2 cm}
\end{table}



\paragraph{NE-efficiency in Reducing the MTD Cost}
We show the efficiency of the game-theoretic solution in reducing the operator's defense cost. 
The simulations are done on a IEEE-14 bus system. The generation cost is assumed to be linear, i.e., $C_i (G_{i,t}) = c_i G_{i,t}.$ The generators' capacities at buses $1,\ 2,\ 3,\ 6,\ 8$ are $G_{\max} = 300,\ 50,\ 30,\ 50,\ 20$~MWs and $c_i = 20,\ 30,\ 40,\ 50,\ 35$~\$/MWh respectively. $\fv_{\max}$ is chosen to be $160$ MWs for link $1,$ and $60$ MWs for all other links.
We consider two scenarios: (1) heavily loaded system, with the load values at Bus 1 to 14 given by $0,\ 21.7,\ 94.2,\ 47.8,\ 7.6,\ 11.2,\ 0,\ 0,\ 29.5,\ 9,\ 3.5,\ 6.1,\ 13.5$, $14.9$ MWs respectively, and (2) lightly loaded system, with the load values at Bus 1 to 6 $0,\ 80,\ 44.2,\ 47.8,\ 30,\ 11.2$~MWs respectively and zero loads at Bus 7 to 14. 
We consider five MTD perturbation strategies for the defender, i.e., $d_1 = \{1\}, d_2 = \{1,3\}, d_3 = \{1,3,5\}, d_4 = \{1,3,5,8\}, d_5 = \{1,3,5,8,9,18,19\}.$ We note that $d_5 = \mathcal{L}_D,$ which protects all the links of the system from CCPA. In each case, we perturb the link reactance by $15 \%$ of their original values. The attacker in turn launches a CCPA by disconnecting one of the links at a time. Under this set-up, we compute the NE solution according to 
EXP3 algorithm, with $\gamma_t = \beta_t = 0$ and $\eta_t = 0.01$.

The evolution of $\bar{p}_{k,t}$ in the two scenarios
are shown in Fig.~\ref{fig:Exp_conv}. First, we have verified that the EXP3 and the Lemke-Howson algorithm (based on complete and perfect game knowledge) lead to the same NE solution, which validates our approach. Then, the convergence rate of EXP3 is also reasonably fast (within $10^4$ iterations). Furthermore, it can be observed that the NE robust solution depends on the system load. While, in the heavily loaded scenario, all the links in $\mathcal{L}_D$ need to be perturbed, in the lightly loaded scenario, it is sufficient to perturb the reactance of link 1 only. The rationale is that in the lightly loaded scenario, only a subset of links need to be protected from physical attacks, since the attacker is unlikely to target the unimportant links (i.e., the links that have very little power flow). 

In Table~\ref{tbl:NE}, we also list the defense cost as the percentage increase in the OPF cost over its optimal value (by solving  \eqref{eqn:OPF_normal}). The NE solution of scenario~2 incurs much lower defense cost, since only a subset of links are perturbed. The above experiments show that the MTD perturbation set depends on the operational state of the system. Also, by exploiting the NE solution, the operator can reduce its defense cost.

\begin{figure}[!t]
  	\centering
  	\includegraphics[width=0.45\textwidth]{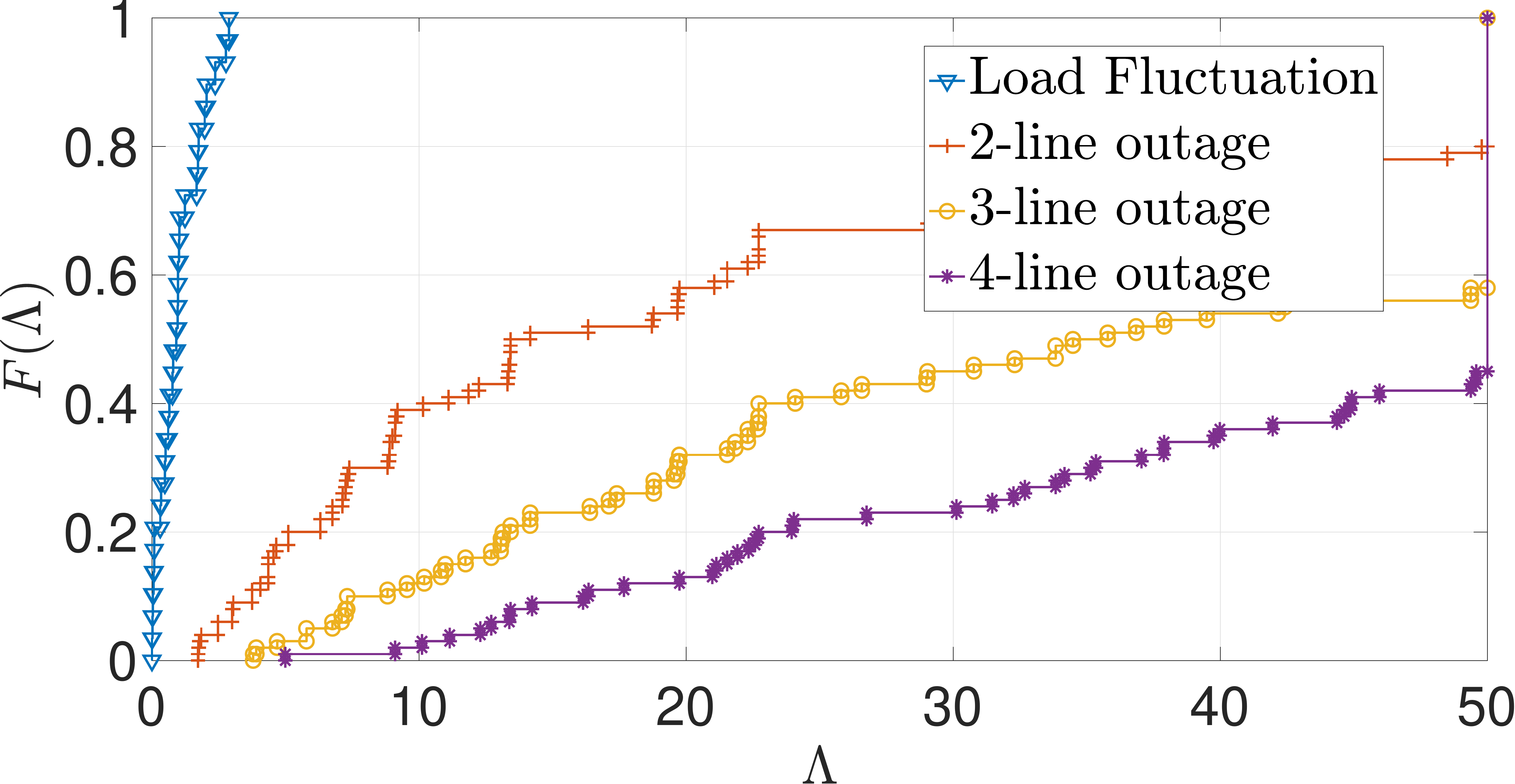}
  	\caption{Cumulative distribution function (CDF) of $\Lambda$ over $100$ combinations of multiple line outages and measurements derived from normal load fluctuations.} 
  	\label{fig:Diff_Mes}
  	\vspace{-0.2 cm}
  \end{figure}

We also examine in the change in the measurements due to CCPA with respect  to the original measurements due to multiple line disconnections. 
We conduct simulations using the IEEE-14 bus system and 
record $\Lambda$ defined as 
$$\Lambda = \max_{i = 1,\dots,M}\frac{ (\zv_p + \Delta H \theta_p)_i - \zv_i }{\zv_i}$$ for $100$ different combinations of multiple line outages ($2-,3-$ and $4-$ simultaneous line outages). We plot the cumulative distribution function (CDF), $F(\Lambda)$ of $\Lambda$ obtained by the $100$ different line outage combinations. We also plot the CDF of the difference is successive measurements due to the normal load changes (the load data is obtained from New York state, available online \cite{NYISO}). We observe that the value of  $\Lambda$ is very high  for almost all combinations of multi-line outages as compared to the fluctuations due to normal load changes, thus verifying that such attacks can be easily detected by the system operator. Thus, in practice, the action set of the attacker required to compute the NE can be limited to single line disconnections only.

\paragraph{Effectiveness of the Proposed MTD under the AC Power Flow Model}
We investigate here the effectiveness of the proposed MTD design in detecting CCPA attacks under an AC power flow model. First note that under AC, the measurements are given by $\zv = \hv(\vv) + \nv,$ where the function $\hv(\vv)$ is a non-linear mapping between the system state (which includes both the voltage magnitudes and the phase angles) and the measurements. 
The new measurements following the physical attack are denoted by $\zv_p,$ where $\zv_p = \hv_p(\vv_p) + \nv.$ Here in  $\hv_p$ and $\vv_p$ are the non-linear measurement function mapping and the bus voltages following the physical attack\footnote{Note that $\hv_p$ reflects the modified system topology after line disconnections.}. Similar to the DC case, we seek a cyber attack that will completely mask the effect of the physical attack. Under the AC power flow model, this will be given by $\av = \zv-\zv_p.$ We compute $\zv_p$ using two methods, (i) fully recomputing the power flows and (ii) by using the LODFs.

To compute the efficacy of our proposed MTD design, we conduct the following experiments. 
First, we compute the MTD reactance perturbation according to Algorithm~1 (i.e., based on the DC power flow model). Then, we compute the attacker's undetectable CCPA vector $\av$ based on the AC power flow (described above) using an outdated system model, and we inject it into the measurements $\zv^{\prime}_p$ (after MTD). We then implement the state estimation and bad data detection based on the AC power flow model (with the new reactance values after MTD) and examine the attack detection rate. The results reported in Fig.~\ref{fig:AC_Detection}. It can be observed that the proposed MTD is effective in detecting CCPAs in an AC power flow model. These experiments show that our MTD approach based on the DC power flow model remains effective in the AC power flow model as well.

At last, we have also recorded the time required to compute $N_A \times N_D$ OPF problems under the DC and AC models in Table~\ref{tbl:Time} involved in the game-theoretic payoff computation in Sec. \ref{sec:Game}. The simulations are conducted  using a Windows PC with 3.2 GHz Intel Core i7 processor and 16 GB RAM using the Matpower simulator.
It can be observed that it only takes a few seconds to tens of minutes to compute the payoffs. 
As we have already explained, the time period between successive MTD perturbations can be reasonably long (e.g., in \cite{LakshDSN2018}, we show that hourly MTD perturbations might be realistic for practical systems). Thus, within this interval, it will be entirely feasible to implement the AC OPF problems to compute the payoffs. 

\begin{figure}[!t]
\centering
\includegraphics[width=0.35\textwidth]{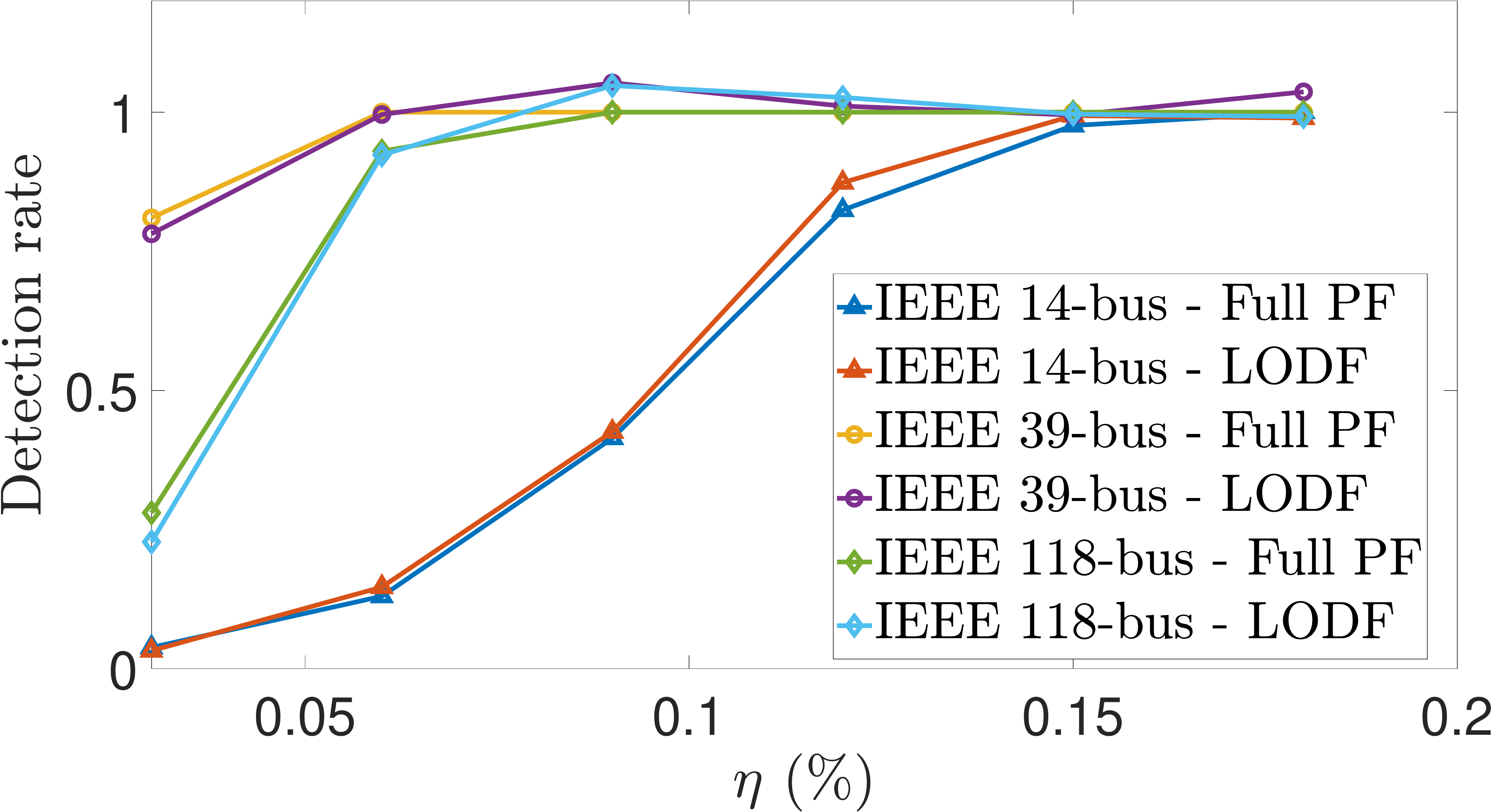}
\caption{{Attack detection probability as a function of the \% change ($\eta$) in the link reactance, for an AC power flow model. The power flows following the physical attack are obtained using two methods (i) full recomputation and (ii) using LODFs.}}
\label{fig:AC_Detection}
\end{figure}

\begin{table}[!t]
 \begin{center}
 \begin{tabular}{|p{1.5 cm} | c | p{1.5 cm}  |  p{1.5 cm} |} 
 \hline
 Bus system & $N_A \times N_D$ & {OPF} computation time in seconds (DC power flow)  & {OPF} computation time in seconds (AC power flow)  \\ [0.5ex] 
 \hline\hline
 IEEE 14-bus system & $20 \times 5 = 100$ & 1.29  & 2.03  \\
 \hline
 IEEE 39-bus system & $41 \times 20 = 820$ & 17.4  & 44.36 \\
 \hline
 IEEE 118-bus system & $186 \times 50 \approx 10000$ & 270  & 600 \\
  [1ex] 
 \hline
\end{tabular}
\end{center}
\caption{Time required to compute $N_A \times N_D$ OPF problems under the DC and AC models involved in the game-theoretic payoff in seconds using the Matpower simulator.}
\label{tbl:Time}
\vspace{-0.2 cm}
\end{table}

\section{Conclusions and Future Work}
\label{sec:Conc}
In this work, we have proposed a novel strategy to detect CCPAs based on MTD and presented MTD design criteria in this context. We have identified the subset of links for D-FACTS device deployment that enables the defender to detect physical attacks against any link in the system. Further, to reduce the operator's defense cost, we have identified the optimal set of links whose reactances must be perturbed at the operational time based on a game-theoretic approach. We showed that the robust solution against a strategic attacker can be computed efficiently exploiting a well-known algorithm in reinforcement learning, which has low complexity and requires little information. 

There are several open research directions that follow from this work. Firstly, a concurrent work \cite{LiuPlacement2020} proposes optimal D-FACTS placement to defend against FDI attacks. Interestingly, their result suggests that the optimal D-FACTS placement to defend against FDI attacks must ensure that the residual graph (obtained by removing the D-FACTS links) has no loops. {This criteria is similar to the one derived in our work, which suggests the possibility of obtaining unified D-FACTS placement algorithm that can defend against both FDI attacks and CCPAs.} {Second, this work focusses on the attack detection problem. An interesting research direction is to exploit MTD to identify the attack, i.e., locate the transmission line disconnected by the attacker and identify the compromised sensors.} 
Third, designing MTD against multi-stage CCPAs and considerations of power grid dynamics via online optimization and learning is an interesting future research direction. Finally, D-FACTS placement algorithm for MTD considering power grid topology reconfigurations is a challenging problem that must be addressed.

\balance
\bibliographystyle{IEEEtran}
\bibliography{IEEEabrv,bibliography}

\end{document}

%% file: macros.tex
\newtheorem{theorem}{Theorem}
\def\LB{\left(}         
\def\RB{\right)}        

\newfont{\bbb}{msbm10 scaled 500}

\newfont{\bb}{msbm10 scaled 1100}

\newcommand{\RR}{\mbox{\bb R}}

\newcommand{\av}{{\bf a}}

\newcommand{\cv}{{\bf c}}
\newcommand{\dv}{{\bf d}}

\newcommand{\fv}{{\bf f}}
\newcommand{\gv}{{\bf g}}
\newcommand{\hv}{{\bf h}}

\newcommand{\lv}{{\bf l}}

\newcommand{\nv}{{\bf n}}

\newcommand{\pv}{{\bf p}}
\newcommand{\qv}{{\bf q}}

\newcommand{\vv}{{\bf v}}
\newcommand{\xv}{{\bf x}}

\newcommand{\zv}{{\bf z}}

\newcommand{\Am}{{\bf A}}
\newcommand{\Bm}{{\bf B}}

\newcommand{\Dm}{{\bf D}}

\newcommand{\Hm}{{\bf H}}

\newcommand{\Wm}{{\bf W}}

\newcommand{\thetav}{\hbox{\boldmath$\theta$}}

\renewcommand{\arg}{{\hbox{arg}}}

\newcommand{\defines}{{\,\,\stackrel{\scriptscriptstyle \bigtriangleup}{=}\,\,}}